\documentclass[useAMS,usenatbib,usegraphicx]{mn2e}
\usepackage{color}
\usepackage{aas_macros}
\usepackage{amsmath}
\usepackage{graphicx}
 \usepackage{multirow}
 \usepackage{float}

\title[]{Supernova-blast waves in wind-blown bubbles, turbulent, and power-law ambient media}
\author[S. Haid]{S. Haid$^{1}$\thanks{E-mail:
haid@ph1.uni-koeln.de}, S. Walch$^{1}$, T. Naab$^{2}$,  D. Seifried$^{1}$, J. Mackey$^{1,3}$ and  A. Gatto$^{2}$\\
$^{1}$I. Physikalisches Institut, Universit\"at zu K\"oln, Z\"ulpicher-Strasse 77, 50937 Cologne, Germany\\
$^{2}$Max-Planck-Insitut f\"ur Astrophysik, Karl-Schwarzschild-Strasse 1, 85741 Garching, Germany\\
$^{3}$Dublin Institute for Advanced Studies, School of Cosmic Physics, 31 Fitzwilliam Place, Dublin 2, Ireland}
\begin{document}

\pagerange{\pageref{firstpage}--\pageref{lastpage}} \pubyear{2015}

\maketitle

\label{firstpage}

\begin{abstract}

Supernova (SN) blast waves inject energy and momentum into the interstellar medium (ISM), control its turbulent multiphase structure and the launching of galactic outflows. Accurate modelling of the blast wave evolution is therefore essential for ISM and galaxy formation simulations. We present an efficient method to compute the input of momentum, thermal energy, and the velocity distribution of the shock-accelerated gas for ambient media (densities of 0.1 $\ge$ $n_{_{0}}$ [$\rm cm^{-3}] \ge$ 100) with uniform (and with stellar wind blown bubbles), power-law, and turbulent (Mach numbers $\mathcal{M}$ from 1 $-$ 100) density distributions. Assuming solar metallicity cooling, the blast wave evolution is followed to the beginning of the momentum conserving snowplough phase. The model recovers previous results for uniform ambient media. The momentum injection in wind-blown bubbles depend on the swept-up mass and the efficiency of cooling, when the blast wave hits the wind shell. For power-law density distributions with $n(r) \sim$ $r^{-2}$ (for $n(r) > n_{_{\rm floor}}$) the amount of momentum injection is solely regulated by the background density $n_{_{\rm floor}}$ and compares to $n_{_{\rm uni}}$ = $n_{_{\rm floor}}$. However, in turbulent ambient media with log-normal density distributions the momentum input can increase by a factor of 2 (compared to the homogeneous case) for high Mach numbers. The average momentum boost can be approximated as $p_{_{\rm turb}}/\mathrm{p_{_{0}}}\ =23.07\, \left(\frac{n_{_{0,\rm turb}}}{1\,{\rm cm}^{-3}}\right)^{-0.12} + 0.82 (\ln(1+b^{2}\mathcal{M}^{2}))^{1.49}\left(\frac{n_{_{0,\rm turb}}}{1\,{\rm cm}^{-3}}\right)^{-1.6}$. The velocity distributions are broad as gas can be accelerated to high velocities in low-density channels. The model values agree with results from recent, computationally expensive, three-dimensional simulations of SN explosions in turbulent media.
\end{abstract}


\section{Introduction}
\label{section1}
 
Supernovae (SN) play a fundamental role in setting the properties of the multi-phase interstellar medium (ISM) (e.g. \citealp{salpeter55,deavillez04,joung06a,kim13,walch15}). They not only enrich the ISM with metals but also inject energy and momentum leading to the dispersal of molecular clouds (MC), the driving of turbulent motions as well as galactic outflows (e.g. \citealp{maclow03,dib06,gent13,girichidis15}). Therefore, SN explosions may locally (and globally) control star formation \citep{agertz13, hennebelle14,iffrig14,walch14a}. Spatially and temporally correlated SNe can interact and drive the expansion of coherent shells, often termed as 'super-bubbles' (e.g. \citealp{mccray87, maclow88, tenoriotagle88,sharma14}). Large-scale super-shells \citep[e.g. Carina Flare;][]{dawson08,palous09, dawson11} may sweep up enough mass to create new MCs, which in turn could spawn new stars and star clusters \citep{elmegreen77, wunsch10, ntormousi11}. On galactic scales SNe might drive fountain flows or even galactic winds (e.g. \citealp{larson74,maclow99, ostriker10,dallavecchia12, hill12,creasey13,girichidis15}). Therefore, SNe might play an important role for regulating the efficiency of galaxy formation and determine galaxy morphology (e.g. \citealp{dekel86, goldbaum11,brook12,aumer13,hopkins14,marinacci14,uebler14}). All of the above conclusions about the impact of SN explosions have been made on the basis of (at the time) computationally expensive numerical simulations with varying degrees of accuracy.  

For a long time the evolution of blast waves has been in the focus of theoretical studies (e.g. \citealp{sedov46,taylor50} and their importance for galactic astrophysics has been realised early on. A key parameter (apart from the explosion energy) determining the fate of a SN remnant (SNR) is the density of the ambient interstellar medium. In numerous analytical studies the evolution of blast waves - also in the presence of cooling - was (mostly) investigated for homogeneous or power-law density distributions \citep{cox72,chevalier76, mckee77, cowie81, cox81, cioffi88, ostriker88, franco94, blondin98}. 

For more realistic density distributions similar to the observed ISM it is more challenging (or even impossible) to make accurate analytical predictions. The ISM is structured and is subject to supersonic turbulent motions, which lead to the observed log-normal shape of the column density probability distribution function \citep[PDF;][]{kainulainen09, schneider11}. Numerical and analytic work confirms a log-normal surface density \citep{maclow03} as well as volume density PDF in isothermal supersonic flows \citep{vazquezsemadeni93, padoan97, padoan97a, kritsuk06, federrath08, ostriker01, walch11b, shetty12, ward14}. In addition, the structure of the ISM around massive stars is strongly affected by the massive stars' ionizing radiation (e.g. \citealp{kesseldeynet03, dale05, gritschneder09, walch12}) and stellar winds (e.g. \citealp{weaver77}). These structural changes affect the impact of SN explosions (e.g. \citealp{rogers13,walch14a,geen15}).  

The efficiency with which energy and  momentum from a SN explosion is transferred to the ambient medium depends on the mean ambient density $n_{_{0}}$ and its turbulent Mach number $\mathcal{M}$. Direct numerical simulations indicate that in dense environments ($n_{_{0, \rm  turb}}$ = 100 $\mathrm{cm^{-3}}$) and low-Mach-number regimes ($\mathcal{M}$ $<$10) the input of momentum is moderate in the presence of cooling \citep{walch14a, kim14} with a momentum transfer of $\sim$ 10 times the initial SN momentum $\mathrm{p_{_{0}}}$ ($\rm p_{_{0}}$ $\sim$ $10^{4} -  3 \times 10^{4}$ $\rm M_{\odot}\, km\,s^{-1}$, in this work $\rm p_{_{0}}$ = 14181 $\rm M_{\odot}\, km\,s^{-1}$), while the momentum input can be $\sim$ 2 times larger for densities $n_{_{0, \rm turb}}<$ 0.1 $\mathrm{cm^{-3}}$. For lower densities, however, the energy and momentum transfer can be significantly higher. Recent numerical simulations have shown that varying assumptions for typical ambient densities of SN explosions can result in very different evolutionary paths of the ISM. In the most extreme case of SN mainly going off in the diffuse phase, the SNRs can interact without significant cooling and the system can go into thermal runaway or start driving a hot outflow \citep{gatto15,girichidis15,li15}.

In cosmological simulations of galaxy formation with typical resolution elements of several hundred parsecs, all the above details - in particular the first phases of blast wave evolution - are unresolved in dense environments, leading to discrepancies between the theoretical expectations and the simulated reality (see e.g. \citealp{schaye15}). In general, this long-known 'over-cooling problem' appears when the main momentum creating stages, the Sedov-Taylor and the pressure driven snowplough phase, stay unresolved and become artifically short \citep{balogh01, stinson06, creasey11, tomassetti15}. The thermal energy is radiated away too quickly and the momentum input is unresolved as too much mass is accelerated to too low velocities \citep{hu15}, in particular if the time step is not reduced accordingly \citep{dallavecchia12, kim14}. The properties of the hot phase within the SNR are also predicted inaccurately and the effect on the global filling factor of the ISM is then biased \citep{mckee77, agertz13,keller15}. A plausible way to overcome these inaccuracies might be the construction of sub-resolution feedback models with information extracted from small-scale resolved numerical simulations of SNRs. However, this computationally expensive process has to cover all the complexity of SNRs and their surroundings \citep{martizzi14,thompson14,  walch14a, kim14}. 

To better understand the evolution of blast waves in the complex ISM we present an efficient 1-dimensional model, based on the thin-shell approach \citep{ostriker88}, to compute the momentum input from SNe for uniform (see Section \ref{section4_1}), radial power-law (see Section \ref{section4_2}), wind-blown bubble (see Section \ref{sec_bubble}) or turbulent environmental density distributions (see Section \ref{section5_1}). In addition to previous studies \citep[e.g.][]{cioffi88, ostriker88} we combine the computation of all blast wave phases and their transitions in a single code using tabulated cooling functions. This way we can cover a wide range of ambient medium parameters. The model is easily customised to different SN scenarios as shown in case of a pre-existing wind bubble or a turbulent environment. We test the code results against recent, highly resolved numerical simulations \citep{martizzi14, thompson14,  walch14a, kim14} and show that we are able to achieve comparable results at almost negligible computational costs. 

The paper is structured as follows. In Section \ref{section2} we discuss the set of equations which govern the evolution of the SNR and the momentum transfer to the ISM. Section \ref{section3} introduces the model which forms the basis for this work. We discuss cases (i) and case (ii) in Section \ref{section4}. In Section \ref{sec_bubble} we use show the momentum input in a wind-blown bubble. In Section \ref{section5} we extend our model to apply it to a turbulent environment and conclude in Section \ref{section7}. 

\section{The evolution of supernova remnants}
\label{section2}

\begin{figure}
\includegraphics[width=0.48\textwidth]{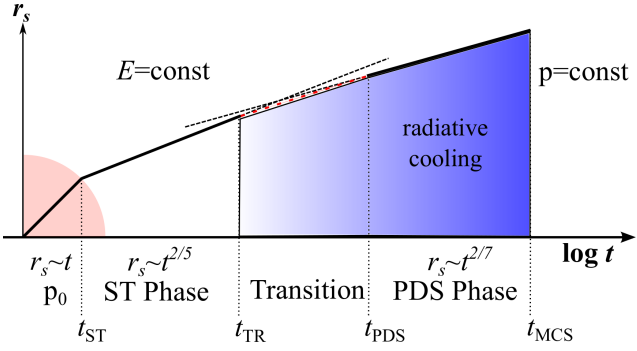}
\caption{Schematic time evolution (times and radius are not to scale) of a SN blast wave radius in a homogeneous environment. $\mathrm{p_{_{0}}}$ is the initial radial momentum of the SN ejecta. The Pre-Sedov-Taylor phase (red) terminates at $t= t_{_{\mathrm{ST}}}$ with the beginning of the energy conserving (non-radiative) Sedov-Taylor (ST) phase ($r{_{\mathrm{S}}} \propto t^{2/5}$). With  radiative losses becoming more important (blue) the blast wave passes through a transition phase $(t=t_{_{\mathrm{TR}}})$ and approaches the fully radiative pressure driven snowplough (PDS) phase at $(t=t_{_{\mathrm{PDS}}})$. The shock radius evolves as $r_{_{\mathrm{S}}} \propto t^{2/7}$ until the momentum conserving snowplough (MCS) phase is reached at $(t=t_{_{\mathrm{MCS}}})$. The swept-up material can only gain radial momentum until the end of the PDS phase.} 
\label{fig:stages}
\end{figure}

When a massive star explodes as a core-collapse SN, gas (typically $\mathrm{\sim 2-5\,M_{\odot}}$) is ejected with supersonic velocities  ($v_{_{\mathrm{eject}}} \mathrm{\sim 6000 - 7000\,km\,s^ {-1}}$; \citealp{blondin98, janka12}), and drives a blast wave into the ISM. The evolution of the blast wave can be characterised by the time evolution $t$ of the shock radius $r_{_{\rm S}}$,
\begin{equation}
\label{eq:powerlaw}
r_{_{\rm S}}\propto t^{\eta},
\end{equation}
where $t$ is the time after the explosion and $\eta$ is the expansion parameter \citep{klein94, cohen98, kushnir10}. It can be separated into five different phases (see Fig. \ref{fig:stages}; \citealp{mckee77, cioffi88, ostriker88, petruk06, li15}). 

\begin{itemize} 
\item \textbf{Pre-Sedov-Taylor (PST) phase:} After the initial explosion the density profile of the ejected gas can be approximated with a steep power-law. In this case the shocked ambient medium decelerates the ejecta. The expansion parameter $\eta$ in this ejecta-dominated phase is smaller than one \citep{chevalier82}. As both shocks merge, the SN ejecta move radially outwards with constant velocity $v_{_{\mathrm{eject}}}$ and sweep up the ambient ISM until the swept-up mass is comparable to the ejecta mass $M_{_{\mathrm{eject}}}$. Part of the kinetic energy of the SN ejecta is converted into heat while the shock wave radius evolves as $r_{_{\mathrm{S}}} \propto t$.   

\item \textbf{Sedov-Taylor (ST) phase:} At the end of the PST phase about 72 per cent of the initial SN energy is converted into thermal energy and the energy conserving ST phase starts at $t=t_{_{\mathrm{ST}}}$ \citep{taylor50, sedov58, mckee77},
 \begin{equation}
t_{_{\mathrm{ST}}}=\left[ r_{_{\rm S,ST}} \left(\frac{\xi E_{_{\rm SN}}}{\rho_{0}}\right)^{-1/5}\right]^{5/2}
 \end{equation}
with the factor $\xi \sim 2$ and the shock radius $r_{_{\mathrm{S,ST}}}$, which can be
computed as 
 \begin{equation}
 \label{eq:initalR}
r_{_{\rm S,ST}}=\left(\frac{3}{4}\frac{M_{_{\rm eject}}}{\pi \rho_{_{0}}} \right)^{1/3}.
\end{equation}

During the energy conserving ST phase the shock evolves adiabatically with $r_{_{\rm S}} \propto
t^{2/5}$ and the radial momentum of the swept-up mass increases.

\item \textbf{Transition (TR) Phase:}  The energy conserving phase ends when the rate-of-change in temperature due to adiabatic expansion is comparable to radiative losses \citep{ostriker88, petruk06}. In this TR phase, starting at $t=t_{_{\mathrm{TR}}}$, the post-shock cooling time $t_{_{\rm cool}}$ becomes comparable to the age of the remnant (see Section \ref{section2_2_2}) 
\begin{equation}
t_{_{\mathrm{TR}}} \sim t_{_{\rm cool}} \rm .
\end{equation}
The radial momentum can still significantly increase. As the shock front decelerates, the faster post-shock gas compresses the shocked material and forms a thin, dense shell at the end of the TR phase \citep{ostriker88, cioffi88}. 

\item \textbf{Pressure driven snowplough (PDS) phase:} At the beginning of the PDS, at $t=t_{_{\mathrm{PDS}}}$, a dense shell has formed behind the radiative shock \citep{falle75}. Typically  $t_{_{\mathrm{PDS}}}$ is a few times $t_{_{\mathrm{TR}}}$ (see Section \ref{section2_2_2}). The further evolution is dominated by radiation. The homogeneous pressure inside the bubble drives the expansion into the low pressure environment \citep{cox72, gaffet83, cioffi88, cohen98}. The shock velocity and further momentum input to the ISM decrease.

\item \textbf{Momentum-conserving snowplough (MCS) phase:} The MCS phase starts at $t=t_{_{\mathrm{MCS}}}$ once the excess thermal energy is radiated away. The momentum of the shell cannot increase any more. Momentum is conserved and inertia becomes the main driver of the further expansion \citep{cioffi88}. We therefore stop and compare our models at $t_{_{\mathrm{MCS}}}$.

\end{itemize}

\subsection{The Ambient Medium}
\label{section2_1}

The structure and the mean density of the ambient medium have a significant influence on the evolution of a blast wave. Here, we consider the general case of a radial power-law density profile \citep{ostriker88}
\begin{equation}
\label{eq:density}
\rho(r) = \rho_{_{0}}Br^{-\omega},
\end{equation}
where  $\rho_{_{\mathrm{0}}}$ is the central density, $\omega$ is the power-law index and $B$ can be used to normalize the radius \citep{truelove99}.

The mass density is related to the number density, $n$, by $\rho = n \mu m_{_{\mathrm{H}}}$, with $m_{_{\mathrm{H}}}$ being the proton mass and the mean molecular weight $\mu$ (ionized gas with $\mu_{_{\rm i}}\ \mathrm{=0.61}$; atomic gas with $\mu_{_{\rm a}}\ \mathrm{=1.27}$).

The total mass of the SNR, $M$, is
\begin{equation}
M(r)=M_{_{\rm eject}}+\frac{4}{3-\omega}\pi  \rho_{_{0}} B  r_{_{\rm S}}^{3 -\omega}\ \  \mathrm{for}\ \omega \neq 3 ,
\label{eq:mass}
\end{equation}
where $M_{_{\rm eject}}$ is the mass of the SN ejecta. The second term corresponds to the swept-up mass. As the PST phase is dominated by the mass of the ejecta, we assume a constant density, $\rho_{_{0}}$ until $t_{_{\mathrm{ST}}}$. In the following we describe in detail our numerical model considering the different phases starting with the ST phase.

\subsubsection{Sedov-Taylor phase}
\label{section2_2_1}
At the beginning of the adiabatic ST phase a certain percentage of the initial kinetic energy has thermalized (approximately 75 per cent in a homogeneous medium). The fraction of kinetic to thermal energy stays constant and the total energy is conserved \citep{chevalier76, cioffi88}.

At $r_{_{\rm S, ST}}$ (Eq. \ref{eq:initalR}) the adiabatic expansion begins with the radial evolution of the shock, described by the Sedov solution \citep{sedov46, newman80, ostriker88, klein94, truelove99, breitschwerdt12},
\begin{equation}
r_{_{\rm S}}(t)=\left(\frac{\xi E}{\rho_{_{0}}B}\right)^{\frac{1}{5-\omega}}t^{\frac{2}{5-\omega}}
\label{eq:initialSTrad}
\end{equation}
with $\xi=(5-\omega)(10-3\omega)/8\pi$ and the expansion parameter $\eta = 2/(5-\omega)$.

The expansion speed can be derived by considering the time derivatives of the shock radius $r_{_{\rm S}}$ in the ST stage \citep{cavaliere76}:
\begin{equation}
\label{eq:STvel}
\frac{d}{dt}(r_{_{\rm S}})=v=\frac{2}{5-\omega}\frac{r_{_{\rm S}}}{t}.
\end{equation}
Here $v$ is the shock velocity. The post-shock velocity $v'$ is
\begin{equation} 
v' = 3/4 v.
\end{equation}

\subsubsection{Transition phase}
\label{section2_2_2}
Between the ST and PDS phases, there is an intermediate period of non-self-similar behaviour which, therefore, cannot be described by a power-law solution as in Eq. \eqref{eq:powerlaw}. We treat the TR phase independently, which allows a more realistic modelling of the SNR \citep[e.g.][]{cioffi88, petruk06}. The description of the ST phase as energy conserving is accurate as long as cooling plays a minor role and the energy loss due to radiation is negligible. 

Following \citet{blondin98} $t _{_{\mathrm{TR}}}$ is defined as the time at which the cooling time is comparable to the age of the remnant. We obtain similar results when the rate of change in temperature of the SNR, $T$, due to the adiabatic expansion becomes comparable to the radiative losses \citep{petruk06}:
\begin{equation}
\label{eq:t_trans}
 \frac{d}{dt_{_{\rm TR}}}\left(T\right)_{\rm exp} \sim \frac{d}{dt_{_{\rm TR}}}\left(T\right)_{\rm cool}.
\end{equation}

During the TR phase the post-shock gas velocity approaches the shock speed \citep{cioffi88},
\begin{equation}
v'=K_{_{01}}\nu_{1}v,
\end{equation}
with the velocity moment, $K_{_{\mathrm{01}}}$, and the fraction $\nu_{1}$ of the shock velocity $v$ (see Eq. \ref{eq:STvel}). 

The velocity moment, $K_{_{\mathrm{01}}}$,  is unity in self-similar blast waves but changes whenever this condition is violated, thus at $t _{_{\mathrm{TR}}}$, $K_{ _{\mathrm{01,TR}}}\ \mathrm{=0.857}$ \citep[][but see also \citealt{ostriker88}, for more details]{cioffi88}. 

 We follow \citet{cioffi88} and assume that the TR phase lasts until
 \begin{equation}
 \label{eq:euler}
 t_{_{\mathrm{TR}}} c = t_{_{\mathrm{PDS}}}
 \end{equation}
where $c$ = $(1+\eta) / (\eta ^{\eta /(1+ \eta)})$ with $\eta = (4(3-\omega)-2\omega)/(5-\omega)$. We follow the approximation by \citet{petruk06} and assume $c$ = 1.83 for the homogeneous medium and $c$ = 1 for $\omega$ = 2. During this period, $\nu_{1}$ changes as
\begin{equation}
\nu_{1}=\frac{3}{4}+0.25\left(\frac{\left(\frac{t}{t_{_{\mathrm{TR}}}}\right)^{2.1}-1}{\left(\frac{1}{c}\right)^{2.1}-1}\right).
\end{equation}

As radiative cooling becomes important, $\nu_{1}$ increases from the ST value of 3/4 to a value of one at $t_{_{\rm PDS}}$.  A thin, dense, radiatively cooling shell forms \citep{gaffet83, ostriker88, cioffi88, petruk06}. 

The large thermal pressure gradient across the shock drives the expansion under the influence of radiative cooling \citep{cioffi88}. We use a set of coupled ordinary differential equations for the further evolution of the SNR starting at $t_{_{\mathrm{TR}}}$, throughout the PDS phase until $t_{_{\mathrm{MCS}}}$.
The time evolution of mean momentum and shock radius then read (see \citet{ostriker88}, their Eq. (2.9) and appendix D): 
\begin{equation}
\label{eq:PDSmv}
\frac{d}{dt}(\bar{p})= \frac{4(3-\omega) \pi}{3} K_{_{\mathrm{pres}}}\bar{P}_{_{\rm th}}r_{_{\rm S}}^{2}
\end{equation}
\begin{equation}
\label{eq:PDSvel}
\frac{d}{dt}(r_{_{\rm S}})=\frac{3}{4r_{_{\rm S}}^{3}\pi \bar{\rho}}\frac{1}{K_{_{\mathrm{01}}}\nu_{1}}(\bar{p})
\end{equation}

where $K_{_{\mathrm{pres}}}$ is the pressure moment and $\bar{P}_{_{\rm th}}$ is the mean thermal pressure within the SNR, 
\begin{equation}
\label{eq:pressure}
\bar{P}_{_{\rm th}}=\frac{E_{_{\mathrm{th}}}}{2\pi r_{_{\rm S}}^{3}},
\end{equation}
which depends on the thermal energy $E_{_{\mathrm{th}}}$ of the SNR changing as
\begin{equation}
\label{eq:energyrate}
\frac{d}{dt}(E_{_{\mathrm{th}}})=-V \Lambda (\bar{T})\bar{n}^{2}.
\end{equation}
$\Lambda$ is the cooling function (see Section \ref{section3}) in a volume $V$ with a mean number density $\bar{n}$ and a mean temperature $\bar{T}$. We consider two volumes, namely that of the shock and the interior. Note that Eq. \ref{eq:energyrate} is used throughout the entire evolution of the SN blast wave from $t_{_{\rm ST}}$ until the end \citep{ostriker88, bisnovatyi95}. During the ST phase almost no thermal energy is radiated away. Internal structures have minor influence compared to the shock and are therefore neglected.

The pressure moment, $K_{_{\mathrm{pres}}}$, can be interpreted as the weighted mean interior pressure of the SNR (see \cite{ostriker88}, Eq. D10a for further details). At the beginning of the TR phase in our SN-model $K_{_{\mathrm{pres, TR}}}$ = 0.932 and approaches $K_{_{\mathrm{pres, PDS}}}$ = 1 \citep{cioffi88, ostriker88, bisnovatyi95}.

\subsubsection{Pressure driven snowplough phase}
\label{section2_2_3}
The PDS is the first fully radiative phase. It starts with the formation of a thin shocked shell, which contains most of the mass of the SNR and encloses a roughly isobaric and hot cavity  \citep{blondin98}. Since we restrict ourselves to one dimension, we neglect instabilities or deviations from spherical geometry \citep{franco94}.

The evolution during the PDS is also described by the equations introduced in Section \ref{section2_2_2} with $K_{_{\mathrm{pres}}}$ = $K_{_{\mathrm{01}}}$ = $\nu_{1}$ = 1. With a dense, uniform, thin shell we can model the flow using a self-similar solution and Eq. \eqref{eq:powerlaw} is valid. As we neglect the influence of the inner parts, the expansion parameter $\eta$ in this case is \citep{ostriker88, gaffet83},
\begin{equation}
\eta= \frac{2}{2+3\gamma-\omega},
\end{equation}
where $\gamma$ = 5/3 is the adiabatic index of a mono-atomic gas.

During the PDS almost all thermal energy is radiated away. The thermal pressure inside the cavity becomes equal to the ambient thermal pressure at $t_{_{\mathrm{MCS}}}$. At this point we stop the calculation of the PDS phase and assume that afterwards the radial momentum stays constant.

\section{The numerical setup}

\label{section3}
\begin{table*}
\small{
\begin{tabular}{|l|c|l|l|l|l|c|}
\hline
Initial SN properties &  &  &  &  &\\ 
\hline
\hline
\multicolumn{2}{c}{SN momentum $\rm p_{_{0}}$ = 14181 $\rm M_{\odot}\,km\,s^{-1}$}  & \multicolumn{2}{c}{SN energy $M_{_{\rm eject}}$ = $10^{51}$ erg} & \multicolumn{2}{c}{Ejecta mass $M_{_{\rm eject}}$ = 2 $\rm M_{\odot}$} \\
\hline 
\hline
Property & Structure & Density & \multicolumn{2}{l}{$\ \ \ \ \ \ $ Turbulence} & Figures \\ 
\hline 
\hline
Uniform media ($\mu_{_{\rm a}}$, $\mu_{_{\rm i}}$) & Homogeneous & $n_{_{0, \rm uni}}$ = 0.1 $-$ 100 $\rm cm^{-3}$ & --- & --- & \ref{fig:evoconst}\\ 
\hline 
Media with density gradient & Power-law & $n_{_{0, \rm power}}$ = 0.1 $-$ 100  $\rm cm^{-3}$ & --- & --- & \ref{fig:evograd}\\ 
 \hline 
Different surrounding media & Power-law & $n_{_{0, \rm power}}$ = 1  $\rm cm^{-3}$ & --- & --- & \ref{fig:evogradover}\\ 
 \hline 
 Different initial densities & Wind-blown bubble & $n_{_{0, \rm uni}}$ = 1 $-$ 1000 $\rm cm^{-3}$   & --- & --- & \ref{fig:evoweaver}\\ 
  \hline 
 Different initial temperatures & Wind-blown bubble & $n_{_{0, \rm uni}}$ = 1 $\rm cm^{-3}$   & --- & --- & \ref{fig:evoweaver_temp}\\ 
 \hline
Example ($\mu_{_{\rm a}}$,  $\mu_{_{\rm i}}$)  & Turbulent & $n_{_{0,\rm turb}}$ = 1  $\rm cm^{-3}$ & $\mathcal{M}$ = 10 & $N_{_{\rm cones}}$ = 12 & \ref{fig:mv3} \\ 
\hline 
Density variation & Turbulent & $n_{_{0,\rm turb}}$ = 1  $\rm cm^{-3}$ & $\mathcal{M}$ = 10 & $N_{_{\rm cones}}$ = 12 $-$ 384 & \ref{fig:healpix} (top)\\ 
\hline 
Momentum variation & Turbulent &$n_{_{0,\rm turb}}$ = 1 $\rm cm^{-3}$ & $\mathcal{M}$ = 10 & $N_{_{\rm cones}}$ = 12 $-$ 384 & \ref{fig:healpix} (bottom)\\ 
\hline 
 Momentum at $t_{_{\rm MCS}}$ & Turbulent & $n_{_{0,\rm turb}}$ = 0.1 $-$ 100   $\rm cm^{-3}$ & $\mathcal{M}$ = 0.1 $-$ 100 & $N_{_{\rm  cones}}$ = 192 & \ref{fig:momdens} \\ 
\hline 
Mass-velocity distribution & Turbulent &$n_{_{0,\rm turb}}$ = 1, 100 $\rm cm^{-3}$ & $\mathcal{M}$ = 1, 10 & $N_{_{\rm cones}}$ = 384 & \ref{fig:massvel} \\ 
\hline 
\end{tabular} 
}
\caption{\textit{Top section:} Initial SN properties for all simulations. \textit{Bottom section:} List of performed simulations. Column 1 gives the considered property, column 2 fixes the density structure of the ambient medium, and column 3 defines the density profile. In column 4 and 5 we give the turbulent Mach number and number of cones used to simulate the turbulent sub-structure of the ambient medium (see section 6). The last column lists the corresponding figures in this paper.}
\label{tab:overview_runs}
\end{table*}

We study the evolution of a single SNR from the ST to the MCS phase by solving the set of ODEs (Eq. \eqref{eq:STvel} and \eqref{eq:PDSmv}, \eqref{eq:PDSvel} together with Eq. \eqref{eq:energyrate}) , based on the thin-shell approach \citep{cioffi88, ostriker88}, described in Section \ref{section2_1} via a fifth-order Runge-Kutta-Fehlberg integration scheme \citep{butcher96} with adaptive step-sizing. This spherically-symmetric, 1-dimensional SN model assumes no instabilities in the shell, no shell perforation or internal structures.  An advantage of the presented SN model is, that we can easily and efficiently calculate the evolution of SNe in a large number of different ambient media.


We assume solar metallicity and we model radiative cooling for $\mathrm{10^{4}\ K< T < 10^{8}\ K}$ using the cooling function by \citet{sutherland93}. For $\mathrm{T < 10^{4}\ K}$ a cooling function by \citet{koyama00, koyama02} is used with 
\begin{equation}
\label{eq:cooling}
\begin{split}
\mathrm{\Lambda}{}={}& \mathrm{ \Gamma\left[ 10^{7}exp\left( \frac{-1.184\times 10^{5}}{T+1000} \right)\right.} \\
&\mathrm{+1.4 \times 10^{-2}\sqrt{T}exp\left.\left(\frac{-92}{T} \right)\right]\ erg\ cm^3\ s^{-1}}
\end{split}
\end{equation}
with a fixed heating rate $\Gamma$ \citep{ koyama02, walch14a},
\begin{equation}
\label{gamma}
\mathrm{\Gamma=2\times 10^{-26}\ erg\ s^{-1}.}
\end{equation}

The SN is initialised at the beginning of the ST phase by adding $10^{51}$ erg of total energy $E_{_{\rm SN}}$ \citep{ostriker88} and 2 M$_{_{\odot}}$ \citep{draine11} of ejecta mass at the initial ST radius, Eq. \eqref{eq:initalR}, corresponding to an initial momentum input of $\rm p_{_{\mathrm{0}}}$ = 14181 $\mathrm{M_{\odot}\,km\,s^{-1}}$.

We run simulations with different combinations of ambient medium densities and density distributions (Eq. \eqref{eq:density}, see Table \ref{tab:overview_runs}). The initial number densities for a uniform distribution $n_{_{0,\rm uni}}$ and the central density of the power-law distribution $n_{_{0,\rm power}}$ vary in a range of 0.1 $-$ 100 $\mathrm{cm^{-3}}$ ($n_{_{0,\rm uni}}$ = $n_{_{0,\rm power}}$ = 0.1, 0.3, 1, 3, 10, 30, 100  $\mathrm{cm^{-3}}$). 

At radii smaller than $R_{_{ST}}$ we assume the density to be homogeneous as the mass of the ejecta dominates the first phase. At larger radii we consider different density distributions (constant, power-law, turbulent) in the ambient medium. For the power-law distribution we assume a density floor, $n_{_{\mathrm{floor}}}$:

\begin{equation}
\label{eq:alldensities}
 n_{_{\rm power}}(r) =  \begin{cases}
n_{_{0, \rm power}} &   \mathrm{for}\ r \leq r_{_{\rm ST}}\\
n_{_{0, \rm power}} \left(\frac{r}{r_{_{\rm ST}}}\right)^{-\omega}&  \mathrm{for}\ r  > r_{_{\rm ST}}\\
 & \ \ \ \mathrm{and}\ n_{_{\rm power}}(r) \geq n_{{\rm floor}} \\
n_{_{\rm floor}} & \mathrm{for}\ r  > r_{_{\rm ST}}\\
 & \ \ \ \mathrm{and}\ n_{_{\rm power}}(r) < n_{{\rm floor}}.
\end{cases}
\end{equation}

Without this lower limit the mean of the ambient density would drop to non-physical values and the sound speed of the ambient medium with a fixed pressure would increase to infinity \citep{chevalier76, cavaliere76, greif11, hennebelle12}. 

A self-consistent treatment of the chemical evolution is not included and it is not possible to consider multiple ionization states of the ambient medium. For simplicity, we choose a neutral environment with solar abundances with $\mu_{_{\rm a}}$ = 1.27. Some studies \citep[e.g.:][]{cioffi88, petruk06} consider the SN environment to be ionized. To compare with these results we rerun the simulations in uniform media and for a turbulent example with $\mu_{_{\rm i}}$ = 0.61 (see Section \ref{section4_1} and Section \ref{section5_1}). 

A simulation is terminated at the beginning of the MCS phase, $t_{_{\rm MCS}}$ (see Section \ref{section2_2_3}), after which the momentum is constant. For all environments we assume an universal ambient pressure, because $P \propto nT \sim \rm const$ \citep{mckee77}. All parameters of the model and the performed simulations are summarized in Table \ref{tab:overview_runs}.

The computational effort to run a single SN depends on the number of time steps. The initial step-size is chosen to be a fraction of the ST time, which depends on the density of the ambient medium. During the computation we use adaptive step-size control.  We compare the local, relative error of the radius and the thermal energy obtained from the applied integration scheme with a global tolerance of $10^{-3}$ at densities of $n_{_{0,\rm uni}}$ $\leq$ 50 $\rm cm^{-3}$ and  $10^{-4}$ for denser environments. In case the local error exceeds the global tolerance the time-step is adjusted. On a single core (clock speed 3.40 GHz) a simulations needs between 4$\times 10^{3}$ ($n_{_{0,\rm uni}}$ = 3 $\rm cm^{-3}$) and 1.3$\times 10^{4}$ ($n_{_{0,\rm uni}}$ = 100 $\rm cm^{-3}$) time-steps, which corresponds to a CPU time of 1.5 s to 6 s.

\section{Blastwave evolution in idealised environments}
\label{section4}

\subsection{Homogeneous density distribution}
\label{section4_1}

We apply our model to follow the evolution of blast waves for a single SN in homogeneous media with densities of $n_{_{\mathrm{uni}}}$ = 0.1 $-$ 100 $\mathrm{cm^{-3}}$, covering the more tenuous ISM up to average densities of MCs. We assume both an ionized with $\mu_{_{\rm i}}$ and a neutral ambient medium with $\mu_{_{\rm a}}$.

\begin{figure}
\includegraphics[width=0.5\textwidth]{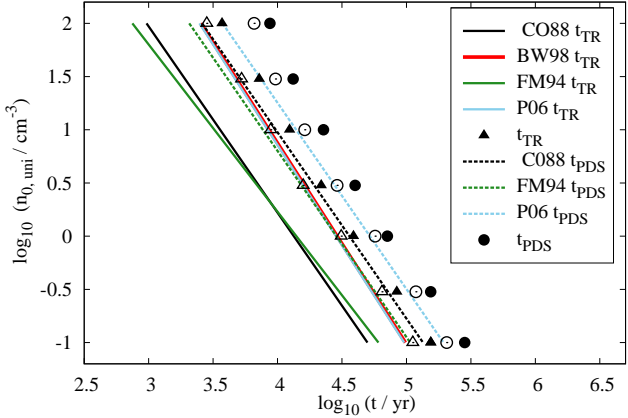} 
\caption{Model predictions for the end of the ST phase $t_{_{\rm TR}}$ (black triangles) and the beginning of the pressure driven snowplough phase $t_{_{\rm PDS}}$  (black circles) in ambient media with different number densities $n_{_{0, \rm uni}}$ and different states of ionization of the ambient gas. Full symbols show the case of a neutral ambient medium with solar abundances ($\mu_{_{\rm a}}$), and open symbols show the case of a fully ionized ambient medium with $\mu_{_{\rm i}}$. Our results are consistent with previous works by \citet[][here BW98]{blondin98} and  \citet[][here P06]{petruk06} but differ significantly from \citet[][here CO88]{cioffi88} and \citet[][here FM94]{franco94} for several reasons (see details in the text).}
\label{fig:transition}
\end{figure}

The transition times $t_{_{\mathrm{TR}}}$ and $t_{_{\mathrm{PDS}}}$ (see Fig. \ref{fig:transition}) of SNe in homogeneous media, obtained in this work, can be fitted with a power-law which depend on the number density $n_{_{0, \rm uni}}$ and mean molecular weight $\mu$ (see Section \ref{section4_1}) :\\
\begin{center}
$t_{_{\mathrm{TR},\mu_{_{\rm a}}}} = 4.15\,(n_{_{0, \rm uni}} / 1\,\mathrm{cm^{-3}})^{-0.53} \times 10^{4}$ yr\\
$t_{_{\mathrm{PDS},\mu_{_{\rm a}}}} = 7.80\,(n_{_{0, \rm uni}} / 1\,\mathrm{cm^{-3}})^{-0.53} \times 10^{4}$ yr\\
$t_{_{\mathrm{TR},\mu_{_{\rm i}}}} = 3.18\,(n_{_{0, \rm uni}} / 1\,\mathrm{cm^{-3}})^{-0.54} \times 10^{4}$ yr\\
$t_{_{\mathrm{PDS},\mu_{_{\rm i}}}} = 5.80\,(n_{_{0, \rm uni}} / 1\,\mathrm{cm^{-3}})^{-0.54} \times 10^{4}$ yr.\\
\end{center}

The definitions for the respective transition times are not unique. Different numerical setups (e.g. \citealt{petruk06}), cooling functions \citep[e.g.][]{cioffi88} and assumptions for the  ambient medium (mean molecular weight in ionized, $\mu_{_{\rm i}}$, or neutral, $\mu_{_{\rm a}}$, media) can lead to different results. Fig. \ref{fig:transition} compares $t_{_{\mathrm{TR}}}$ and $t_{_{\mathrm{PDS}}}$ from previous works \citep{cioffi88, franco94, blondin98, petruk06} to values obtained from this work (black triangles, black circles) in uniform ambient media with number densities between 0.1 $\mathrm{cm^{-3}}$ and 100 $\mathrm{cm^{-3}}$.

Our results are consistent with previous studies by \citet{blondin98} and \citet{petruk06} assuming the ambient medium to be ionized (open symbols). The differences in low density environments are less than 10 per cent. Only at $n_{_{0, \rm uni}}$ = 100  $\mathrm{cm^{-3}}$ the values differ by $\sim$ 40 per cent. In models with a neutral medium (full symbols), $t_{_{\mathrm{TR}}}$ and $t_{_{\mathrm{PDS}}}$ are significantly shifted to later times.  \citet{cioffi88} and \citet{franco94} use different setups and show no agreement with the findings of all other authors.
For a detailed comparison of important times in the evolution of SNRs we refer to \citet{kim14} and \citet{petruk06}. \\

\begin{figure*}
\begin{tabular}{|c|c|}
\begin{minipage}[t]{0.5\textwidth}
\includegraphics[width=1.\textwidth]{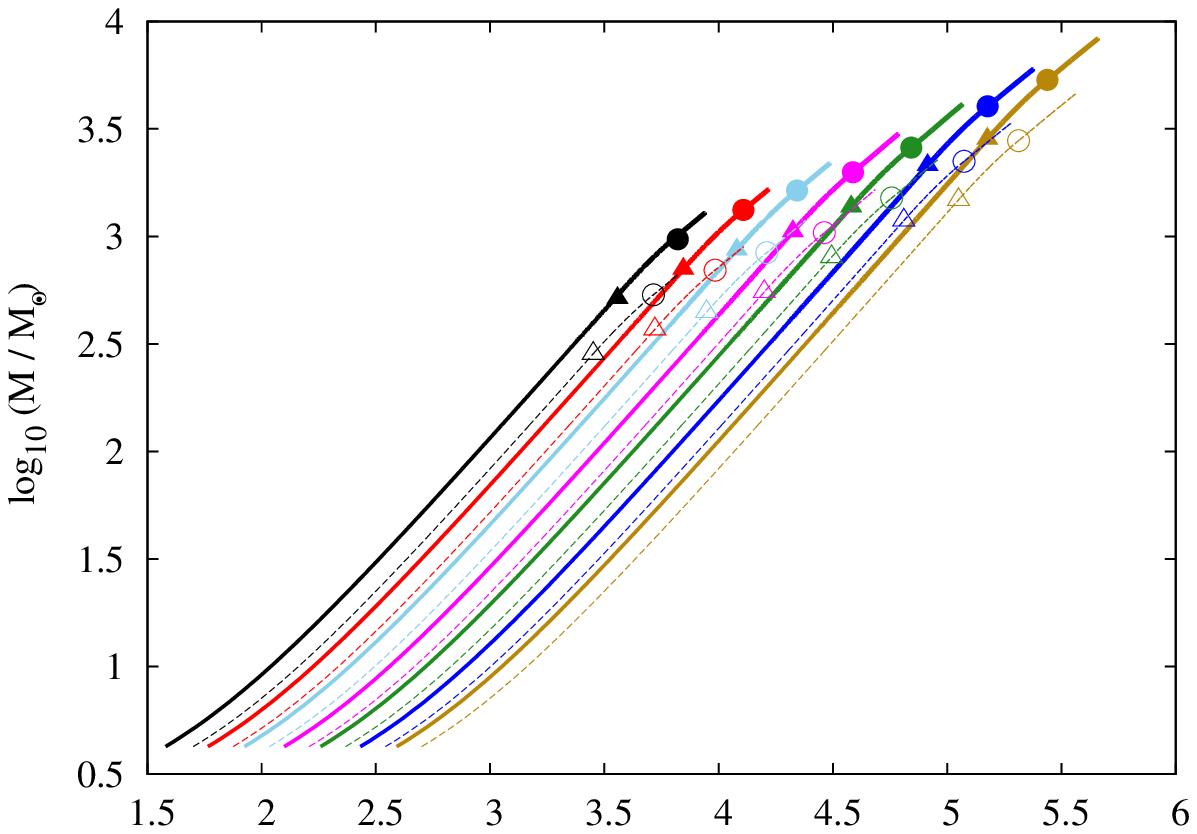}
\end{minipage} & 
\begin{minipage}[t]{0.5\textwidth}
\includegraphics[width=1.\textwidth]{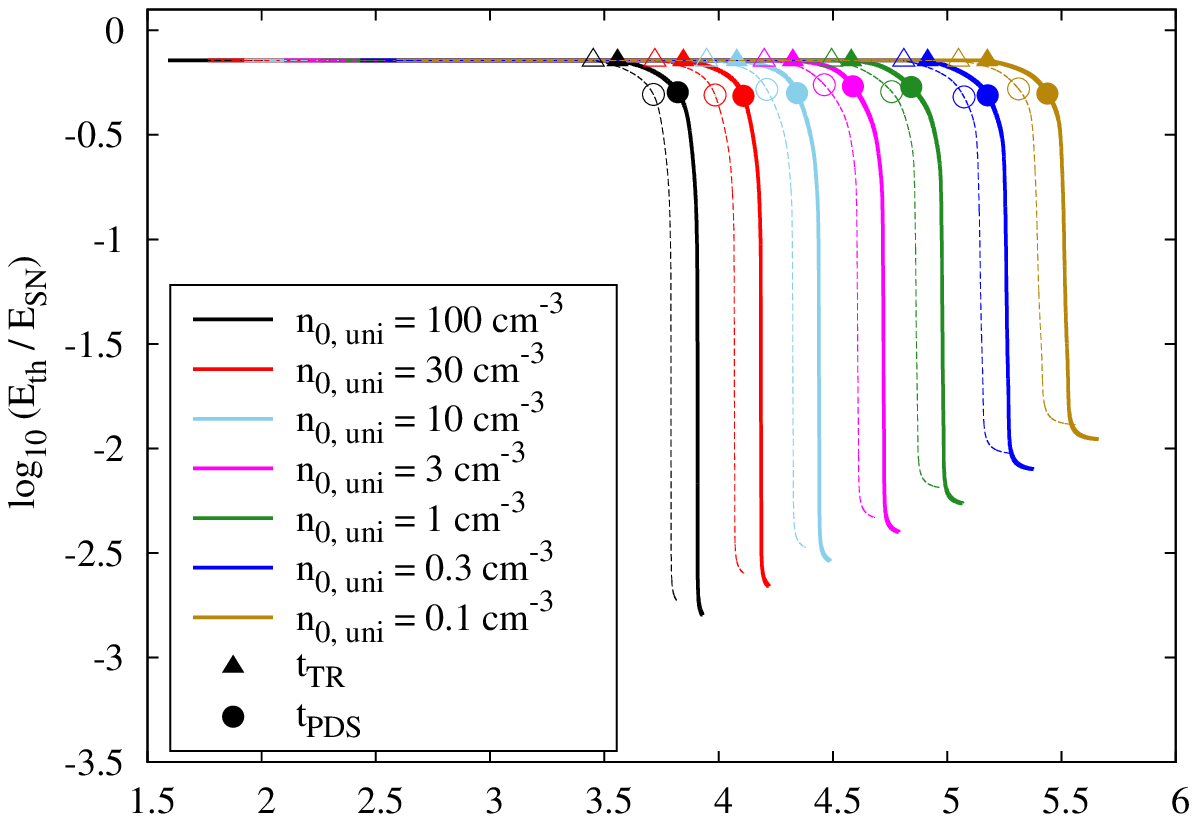}
\end{minipage} \\
\begin{minipage}[t]{0.5\textwidth}
\includegraphics[width=1.\textwidth]{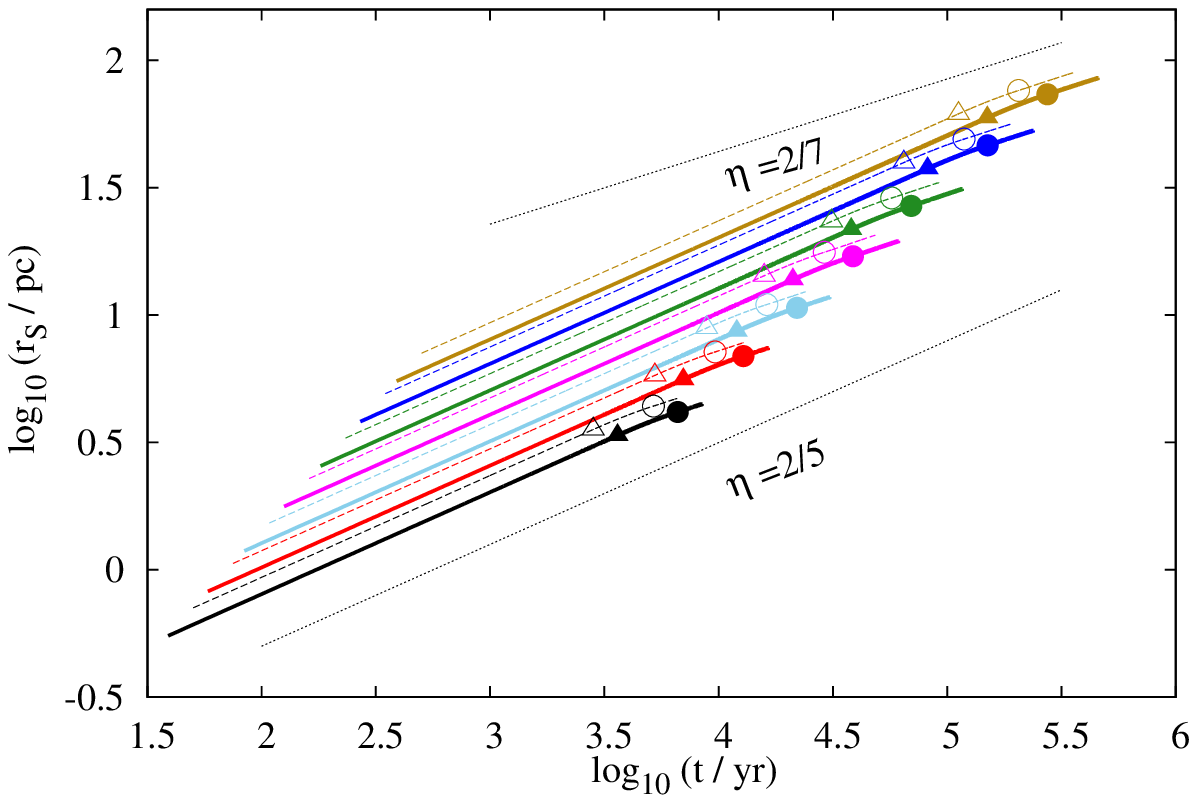}
\end{minipage} & 
\begin{minipage}[t]{0.5\textwidth}
\includegraphics[width=1.\textwidth]{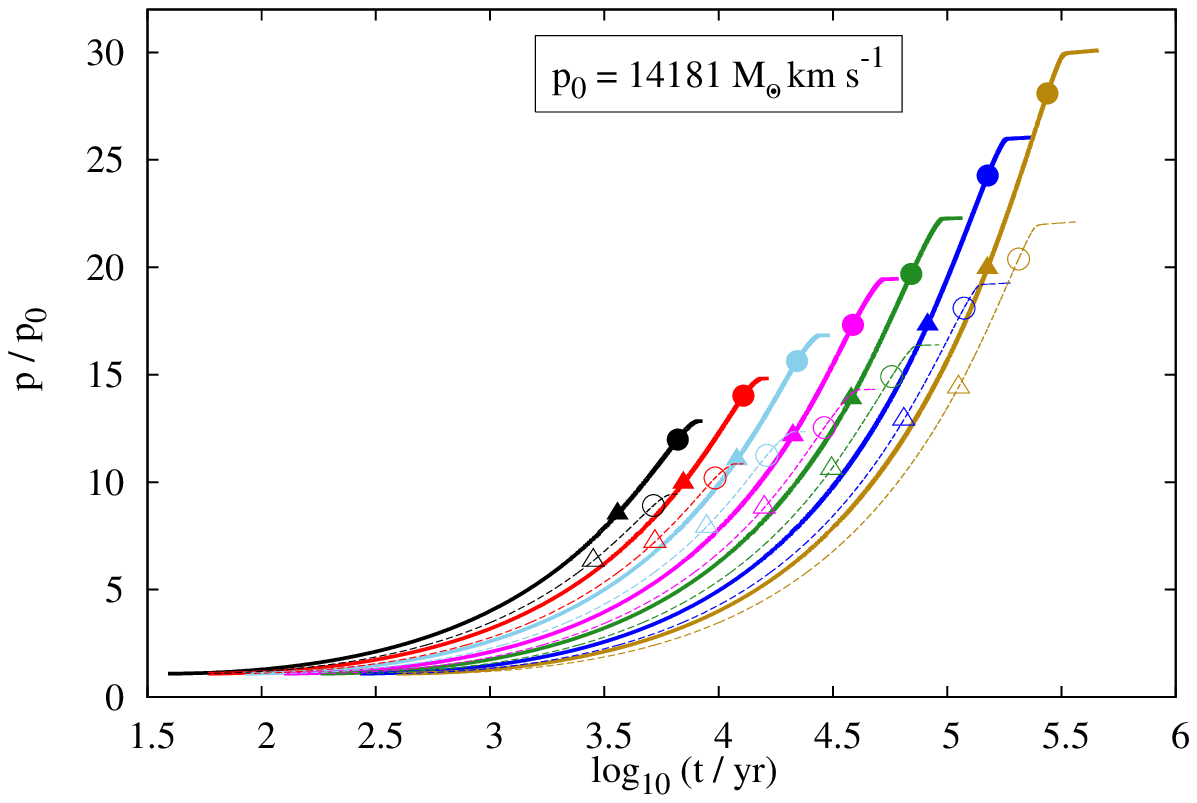}
\end{minipage} \\
\end{tabular}
\caption{Time evolution of SNRs in homogeneous ambient media with densities in the range of 0.1 $-$ 100 $\mathrm{cm^{-3}}$. Triangles indicate the beginning of the TR phase (end of ST phase) at $t_{_{\rm TR}}$, circles the onset of PDS. Open symbols and dashed lines show the corresponding simulation in ionized ambient media. {\it Top left panel:} Time evolution of the swept-up mass. {\it Top right panel:} Evolution of the normalized thermal energy. The energy losses are highest and most rapid for the densest environments. {\it Bottom left panel:} Evolution of the shell radius. The shock radius increases within low density ambient media (up to 85 pc at $n_{_{0,\rm uni}}$ = 0.1 $\mathrm{cm^{-3}}$. Black, dashed lines indicate slopes of 2/5 during t $<$ $t_{_{\rm TR}}$ and 2/7 during t $>$ $t_{_{\rm PDS}}$. {\it Bottom right panel:} Evolution of the momentum input normalized to the initial SN momentum $\rm p_{_{0}}$.} 
\label{fig:evoconst} 
\end{figure*}

In Fig. \ref{fig:evoconst}, top left panel, we show the evolution of the swept-up mass of the SNR. Initially it is dominated by the ejecta mass. The swept-up mass increases rapidly during the ST phase. The final swept-up mass, $M_{_{tot}}$, is $\sim$ 1290 [660] $\mathrm{M_{\odot}}$ in dense environments increasing up to about 8870 [4590] $\mathrm{M_{\odot}}$ in an ambient medium with $n_{_{0,\rm uni}}$ = 0.1 $\mathrm{cm^{-3}}$. This significant increase is a consequence of a 30 times longer evolution in lower-density environments. It will be discussed in more detail in Section \ref{section5_3}.

In Fig. \ref{fig:evoconst} (top right panel) we show the evolution of the thermal energy starting from the ST phase (71.7 per cent of the initial SN energy) until the onset of the MCS phase (end of lines). Here and in all following plots, the beginning of the TR phase is indicated by triangles and the onset of the PDS phase by circles. Filled symbols and thick solid lines show the results for a neutral ambient medium. The open triangles, circles, and dashed lines correspond to the same models assuming an ionised ambient medium. Hereafter, the values for ionised ambient media are given within square brackets.

As expected, for the highest density ($n_{_{0,\rm uni}}$ = 100 $\mathrm{cm^{-3}}$, black line) the ST phase terminates already after 3.6 [2.8] kyr while for the lowest density ($n_{_{0,\rm uni}}$ = 0.1 $\mathrm{cm^{-3}}$, dark yellow line) the ST lasts until 150 [112] kyr. 

As the density of the shell increases, the post-shock gas starts to radiate. At $t_{_{\rm TR}}$ the thermal energy drops significantly at much earlier times for $n_{_{0,\rm uni}}$ = 100 $\mathrm{cm^{-3}}$ than for $n_{_{0,\rm uni}}$ = 0.1 $\mathrm{cm^{-3}}$. For all densities the PDS phase starts at about 1.8 $t_{_{\rm TR}}$. For high densities ($n_{_{0,\rm uni}}$ = 100 $\mathrm{cm^{-3}}$) the PDS phase of 1.9 [1.4] kyr is short compared to 185 [159] kyr in an ambient density of $n_{_{0,\rm uni}}$ = 0.1 $\mathrm{cm^{-3}}$. The bubble stays over-pressured and drives the evolution throughout the PDS stage. Cooling becomes inefficient (the curves flatten toward the end of the evolution) as the temperature of the SNR drops below $\mathrm{10^{4}}$ K \citep[][see Eq. \ref{gamma}]{koyama02, sutherland93}. 

The time evolution of the shell radius is shown in the bottom left panel of Fig. \ref{fig:evoconst}. For all densities the radius evolves as $r_{_{\rm S}} \propto t^{\eta}$ with $\eta = 2/5$ in the ST phase. At $t=t_{_{\rm TR}}$, $\eta$ shifts towards 2/7 and the SNR enters the PDS stage. For the highest density the shell expands to a radius of 3.4 [3.6] pc during the ST and to 4.2 [4.4] pc in the PDS phase. For the lowest density the TR radius is about 59.5 [61.6] pc expanding to 73.5 [76.2] pc in the transition phase and finally reaches 85.3 [90.0] pc at the end of the PDS. The final expansion radius significantly decreases from low to high density environments, because the cooling of the shell occurs earlier and therefore the interior pressure drops more rapidly in denser media. 

\begin{figure*}
\includegraphics[width=0.8\textwidth]{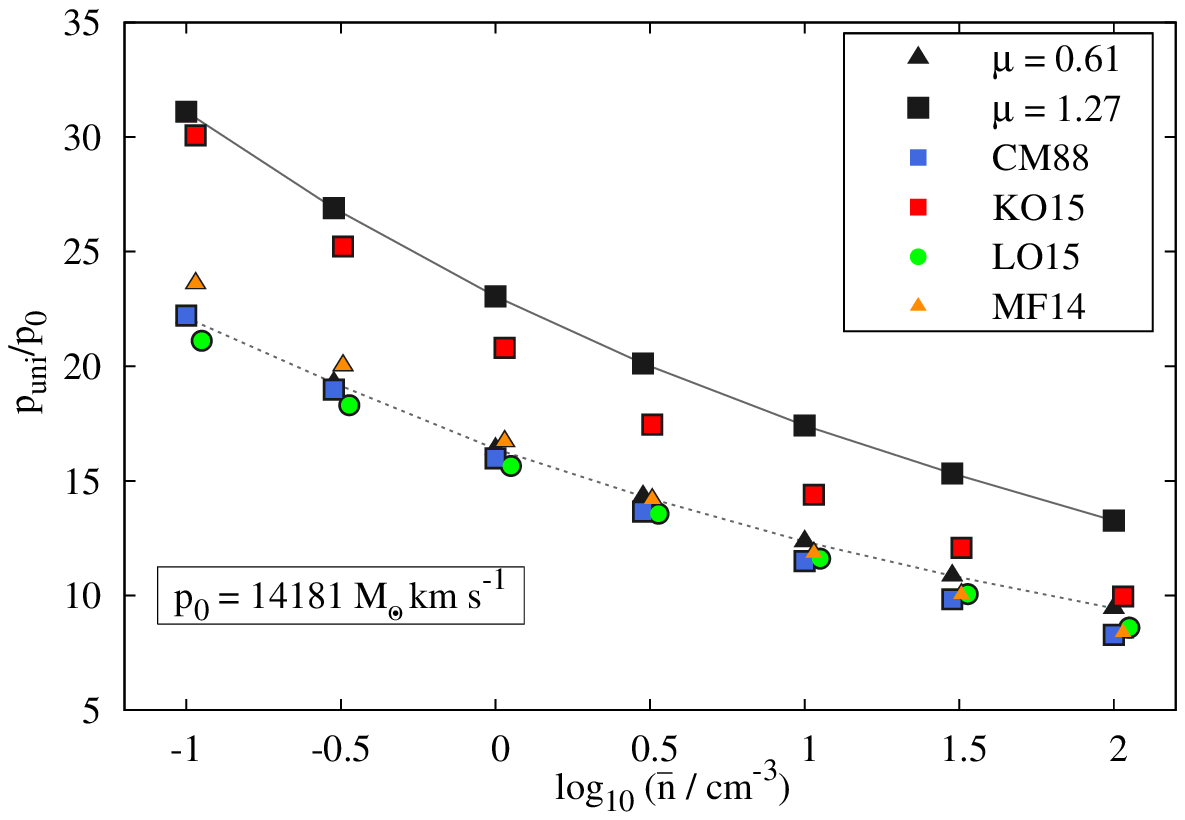}
\caption{Final (at $t_{_{\mathrm{MCS}}}$) radial momentum input $p_{_{\rm uni}}$ in homogeneous medium with densities in the range of $n_{_{0,\rm uni}}$ = 0.1 $-$ 100 $\mathrm{cm^{-3}}$. For comparison we add recent numerical simulations  of SNe in homogeneous media (coloured symbols) from \citet[][here KO15, red squares]{kim14}, \citet[][here MF14, orange triangles]{martizzi14} , \citet[][here CM88, blue squares]{cioffi88} and \citet[][here LO15, green circles]{li15}.} 
\label{fig:evoconst_iter} 
\end{figure*}

In the bottom right panel of Fig. \ref{fig:evoconst} we show the corresponding evolution of the radial shell momentum. During the ST phase the SN momentum increases significantly from $\rm{p_{_{0}}}$ $\approx$ 1.4 $\times 10^{4}\,\mathrm{M_\odot\,km\,s^{-1}}$ by a factor of $\sim$ 8 [6]  for $n_{_{0,\rm uni}}$ = 100 $\mathrm{cm^{-3}}$ and up to a factor 20 [14] at $n_{_{0,\rm uni}}$ = 0.1 $\mathrm{cm^{-3}}$. The following transition phase further increases the momentum by $\sim$ 40 per cent with respect to the ST values. At the beginning of the MCS phase the shell momentum varies between 13.4 [9.3] $\rm{p_{_{0}}}$ for the highest density and 30.9 [21.3] $\rm{p_{_{0}}}$ for an ambient density of 0.1 $\mathrm{cm^{-3}}$. However the momentum increase during the PDS, is almost negligible because the pressure inside the SNR is lowered to values similar to the ambient pressure (see Section \ref{section3}). Within a high density environment ($n_{_{0,\rm uni}}$ = 100 $\mathrm{cm^{-3}}$) the increase is only 0.9 $\rm{p_{_{0}}}$. The final radial momentum converges as the temperature inside the SNR drops. Shortly before the onset of the MCS phase a final plateau forms. The temperature has dropped below 10$^{4}$ K and the photoelectric heating starts to compensate the radiative cooling \citep{koyama02}.
\\
\\
In Fig. \ref{fig:evoconst_iter} we compare the final momenta in a density range of $n_{_{0,\rm uni}}$ = 0.1 $-$ 100 $\mathrm{cm^{-3}}$  from our model with recent numerical simulations \citep{kim14, li15, martizzi14} and with previous works \citep{cioffi88}. We show the results for atomic (full black squares) and ionized media (open black squares). The SN model in an ionized medium with a density of $n_{_{0,\rm uni}}$ = 1 $\mathrm{cm^{-3}}$ has a radial momentum input of $\mathrm{2.3 \times 10^{5}\,M_{\odot}\,km\,s^{-1}}$, which is in good agreement with $\mathrm{2.17 \times 10^{5}\,M_{\odot}\,km\,s^{-1}}$ found by \citet{kim14} with $\mathrm{2.66 \times 10^{5}\,M_{\odot}\,km\,s^{-1}}$ by \citet{li15} and the semi-analytic solution by \citet{cioffi88}.

For neutral and ionised gas the final momentum input is
\begin{center}
 $p_{\mu_{_{\rm a}}}\ =22.44\, (n_{_{0, \rm uni}} / 1\,\mathrm{cm^{-3}})^{-0.12}\ \mathrm{p_{_{0}}} $\\
 $p_{\mu_{_{\rm i}}}\ =16.52\, (n_{_{0, \rm uni}} / 1\,\mathrm{cm^{-3}})^{-0.12}\ \mathrm{p_{_{0}}} $,\\
\end{center}
respectively. Numerical simulations by \citet{kim14} find a lower factor of 19.75 and an exponent of -0.16.

\subsection{Power-law density distribution}
\label{section4_2}

\begin{figure*}
\begin{tabular}{|c|c|}
\begin{minipage}[t]{0.5\textwidth}
\includegraphics[width=1.\textwidth]{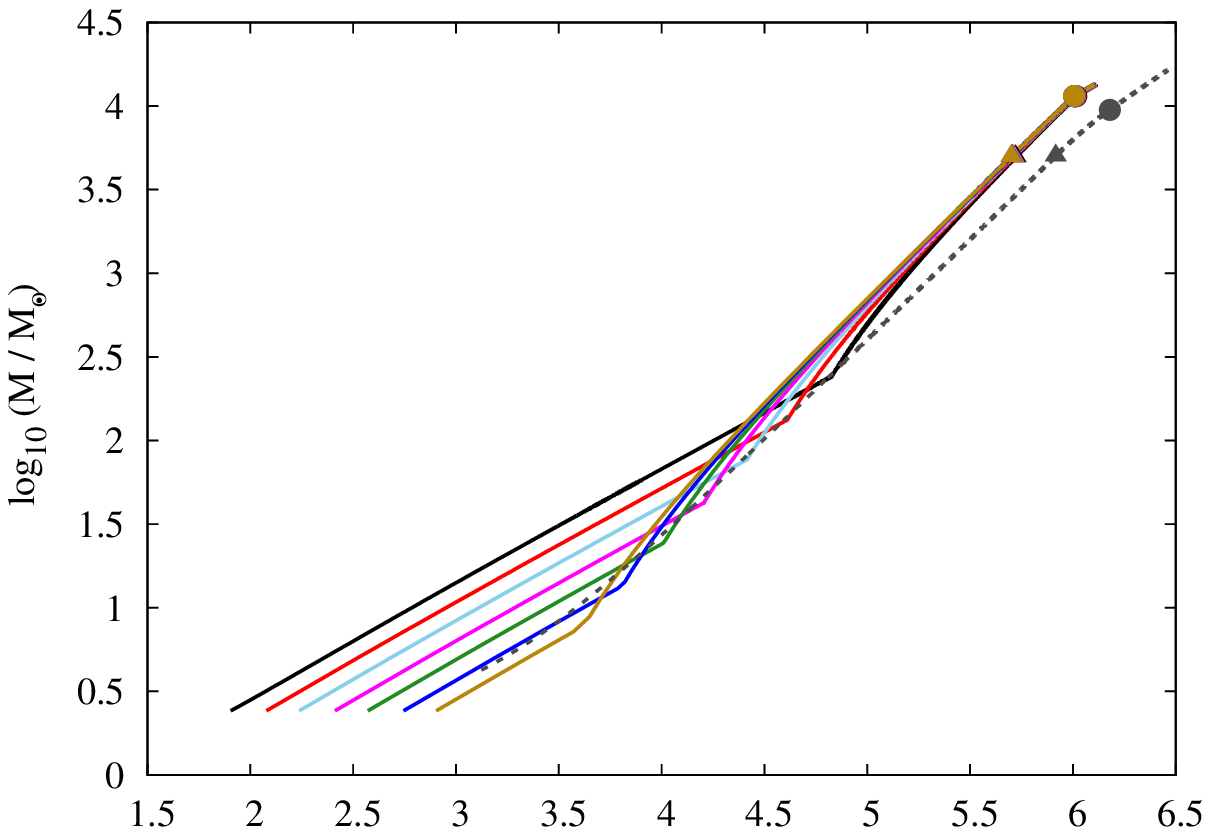}
\end{minipage} & 
\begin{minipage}[t]{0.5\textwidth}
\includegraphics[width=1.\textwidth]{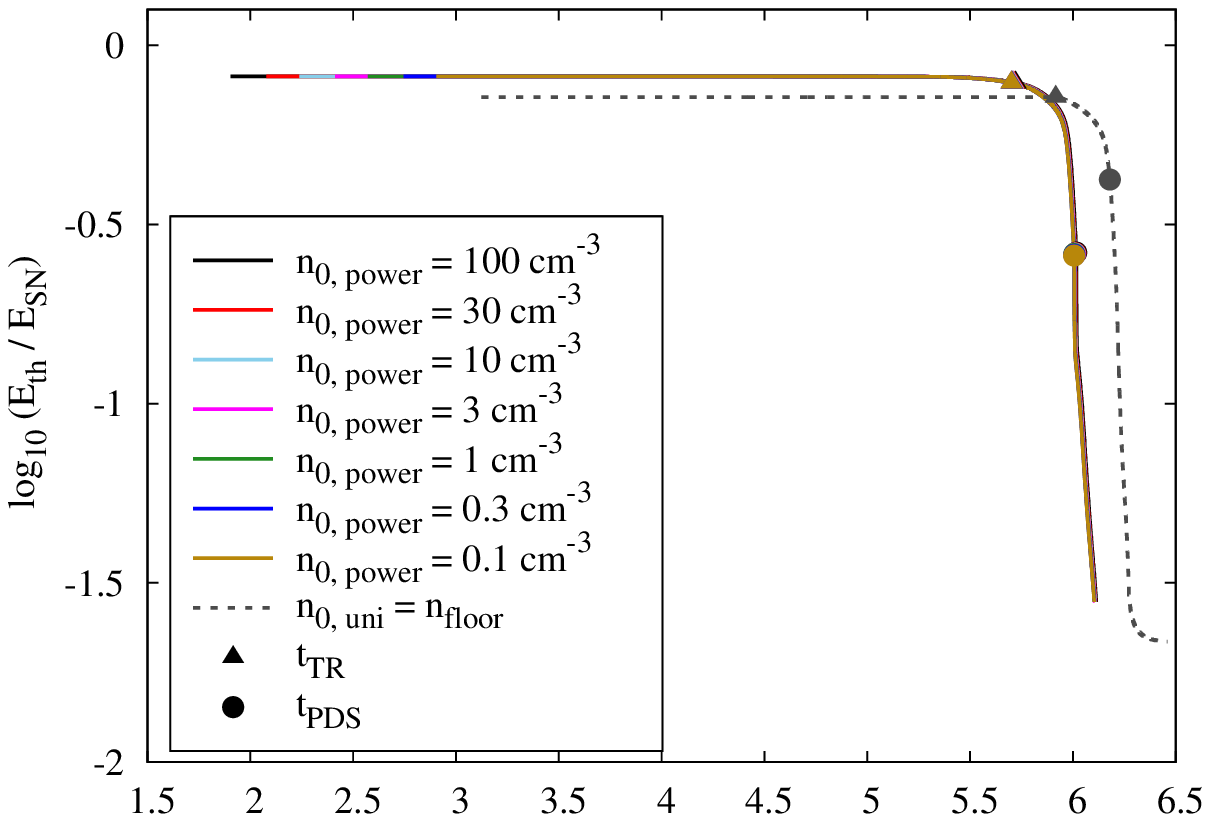}
\end{minipage} \\
\begin{minipage}[t]{0.5\textwidth}
\includegraphics[width=1.\textwidth]{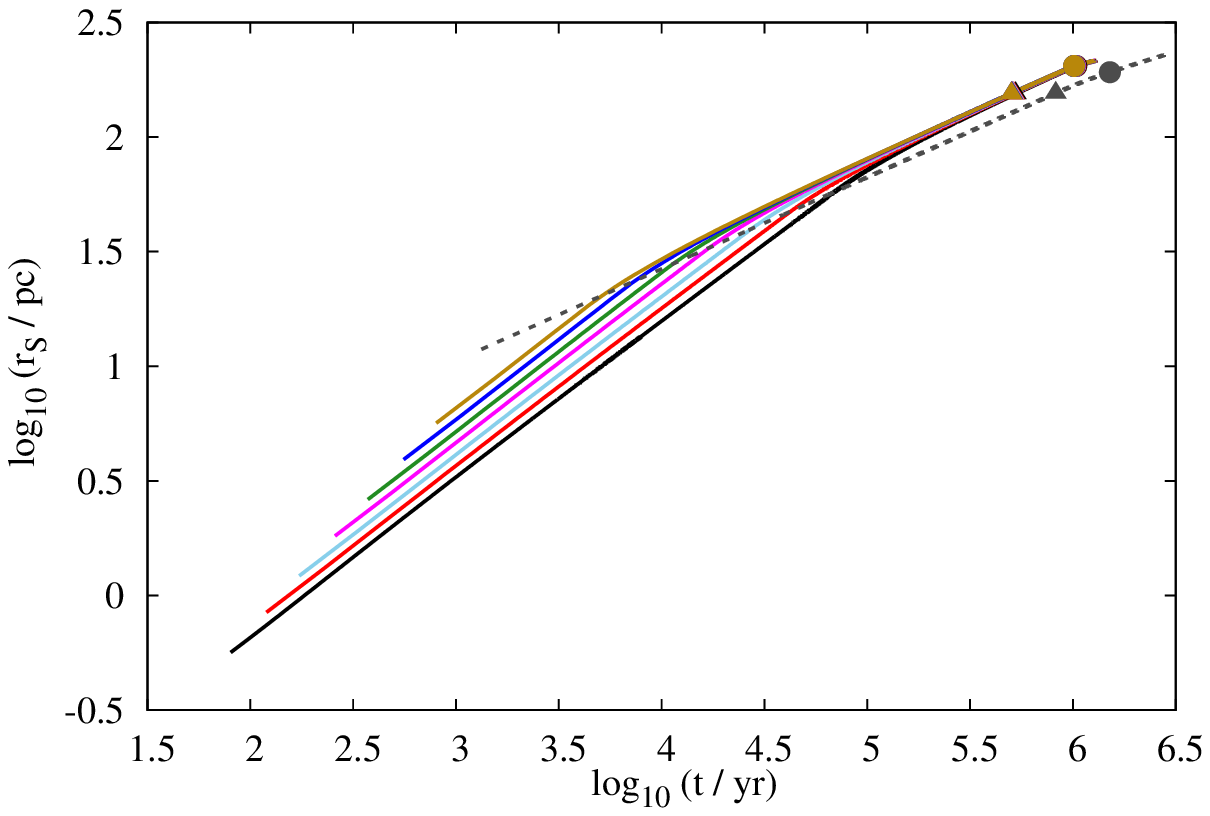}
\end{minipage} & 
\begin{minipage}[t]{0.5\textwidth}
\includegraphics[width=1.\textwidth]{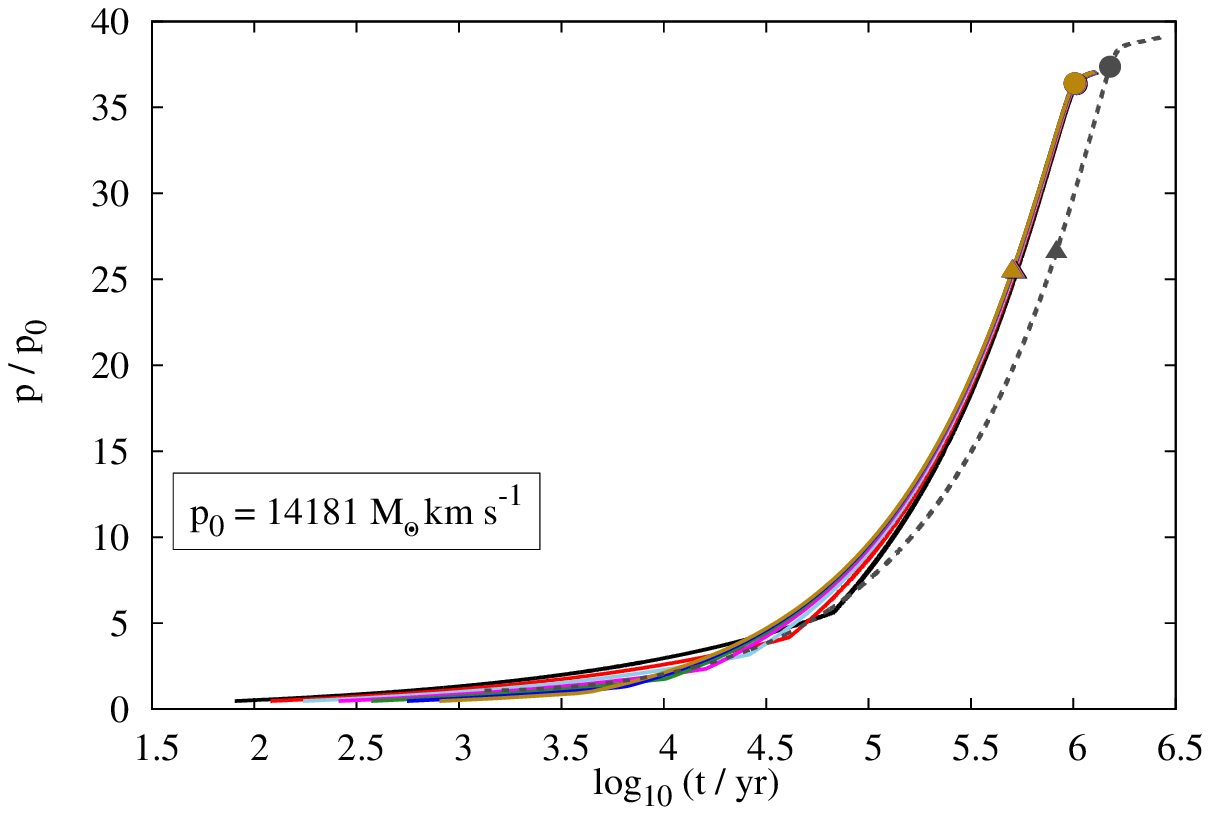}
\end{minipage} \\
\end{tabular}
\caption{Time evolution of SNRs in ambient media with a power-law density distribution and central densities in the range of 0.1 $-$ 100 $\mathrm{cm^{-3}}$ (lines with different colors as indicated in the legend) and a density floor of 0.01  $\mathrm{cm^{-3}}$. Triangles indicate the beginning of the TR phase, $t_{_{\rm TR}}$, circles the onset of the PDS phase. The SN expansion into a homogeneous medium (grey, dashed line) with an ambient density equal to the floor density is shown. It is shifted to later later times by a factor of $10^{0.2}\sim1.59$, because it lies on top of the other lines. {\it Top left panel:} Mass increase during the evolution up to a collective mass of some $10^{3.5} \rm M_{\odot}$. {\it Top right panel:} Evolution of the normalized thermal energy. {\it Bottom left panel:}  Evolution of the shell radius. {\it Bottom right panel:} Evolution of the momentum input normalized to the initial SN momentum $\rm p_{_{0}}$.} 
\label{fig:evograd} 
\end{figure*}

We now assume a power-law ambient medium density distribution following Eq. \eqref{eq:alldensities} with $\omega$ = 2. We vary $n_{_{0,\rm power}}$ = 0.1 $-$ 100 $\mathrm{cm^{-3}}$ \citep{weaver77, band88}.

In the top left panel of Fig. \ref{fig:evograd} we show the corresponding evolution of the swept-up mass. We find two distinct regimes for the mass evolution. Where the ambient density distribution follows a power-law with $M$ $\propto$ $t^{1.95}$. In this medium and a high density ($n_{_{0,\rm power}}$ = 100 $\mathrm{cm^{-3}}$) $\sim$ 155 $\rm M_{\odot}$  is swept-up compared to 6 $\rm M_{\odot}$ for $n_{_{0,\rm power}}$ = 0.1 $\mathrm{cm^{-3}}$. Once the uniform density floor is reached, the swept-up mass is quickly dominated by the surrounding uniform medium with $n_{_{\rm floor}}$. Independent of $n_{_{0,\rm power}}$ the swept-up mass is $\sim$ 5000 $\rm M_{\odot}$ at $t_{_{\rm TR}}$  and 1.3 $\times 10^{4}$ $\rm M_{\odot}$ at $t_{_{\rm MCS}}$. Compared to the uniform ambient medium with $n_{_{0,\rm uni}}$ = 0.01 $\mathrm{cm^{-3}}$, the total swept-up mass in the power-law distribution is $\sim$ 20 per cent smaller. The expansion proceeds shorter in time and expansion in the latter case because slightly less momentum is created during the evolution.

In Fig. \ref{fig:evograd} (top right panel) we show the evolution of the thermal energy normalized to the initial SN energy. The initial thermal energy is 0.82 $E_{_{\rm SN}}$ (results from Eq. \eqref{eq:initialSTrad} and the momentum at  $t_{_{\rm ST}}$). Starting with energy conservation during the ST phase, thermal energy is radiated away at the same $t_{_{\rm TR}}$ (triangles, $t_{_{\rm TR}}\sim$ 510 kyr) independent of the profile density. The thermal energy drops significantly during the PDS phase (circles, $t_{_{\rm PDS}}\sim$ 1 Myr) to 0.26 $E_{_{\rm SN}}$. For all central densities the thermal energy is lost only within the last $\sim$ 300 kyr of the simulation ($t_{_{\rm MCS}}\sim$ 1.2 Myr). For comparison, the thermal energy retained at $t_{_{\rm PDS}}$ in a uniform ambient medium with $n_{_{0,\rm power}}$ = 0.01 $\mathrm{cm^{-3}}$ is 0.4 $E_{_{\rm SN}}$.

The time evolution of the shell radius is shown in Fig. \ref{fig:evograd} (bottom left panel). For all densities the radius evolves with an expansion parameter $\eta \sim 2/(5 - \omega)$ in the ST phase turning to $\eta \sim$ 2/7 as it reaches the PDS phase within the homogeneous medium. For the highest central density ($n_{_{0,\rm power}}$ = 100 $\mathrm{cm^{-3}}$) the shell expands to 155 pc during the ST phase. At $t_{_{\rm PDS}}$ the radius is 204 pc and finally the shell has expanded to 215 pc. These values are almost independent of the central density and are more comparable to the expansion radius of a homogeneous ambient medium with $n_{_{0,\rm power}}$ = 0.01 $\mathrm{cm^{-3}}$, which expands to 230 pc.

The radial momentum (Fig. \ref{fig:evograd}; bottom right panel) depends, among others, on the swept-up mass, which couples the thermal energy to the ambient medium. In a power-law medium, where $n(r)$ decreases rapidly the mass of the SN ejecta dominates the initial evolution (Fig. \ref{fig:evograd}; bottom right panel). The momentum increases between 2.4 $\rm{p_{_{0}}}$ ($n_{_{0,\rm power}}$ = 0.1 $\mathrm{cm^{-3}}$) and 5.1 $\rm{p_{_{0}}}$ ($n_{_{0,\rm power}}$ = 100 $\mathrm{cm^{-3}}$) before $n(r)$ = $n_{_{\rm floor}}$ is reached. From this point onwards, the momentum increases more rapidly. At $t_{_{\rm TR}}$ all simulations converge to a common value of $\sim$ 25.3 $\rm{p_{_{0}}}$, increase to 36.3 $\rm{p_{_{0}}}$ at  $t_{_{\rm PDS}}$  and finally to 37.0 $\rm{p_{_{0}}}$. For comparison, the momentum in a homogeneous medium with $n_{_{0,\rm uni}}$ = 0.01 $\mathrm{cm^{-3}}$ at $t_{_{\rm TR}}$ is 26.6 $\rm{p_{_{0}}}$ and 39.0 $\rm{p_{_{0}}}$ at $t_{_{\rm MCS}}$. \\

\begin{figure*}
\begin{tabular}{|c|c|}
\begin{minipage}[t]{0.5\textwidth}
\includegraphics[width=1.\textwidth]{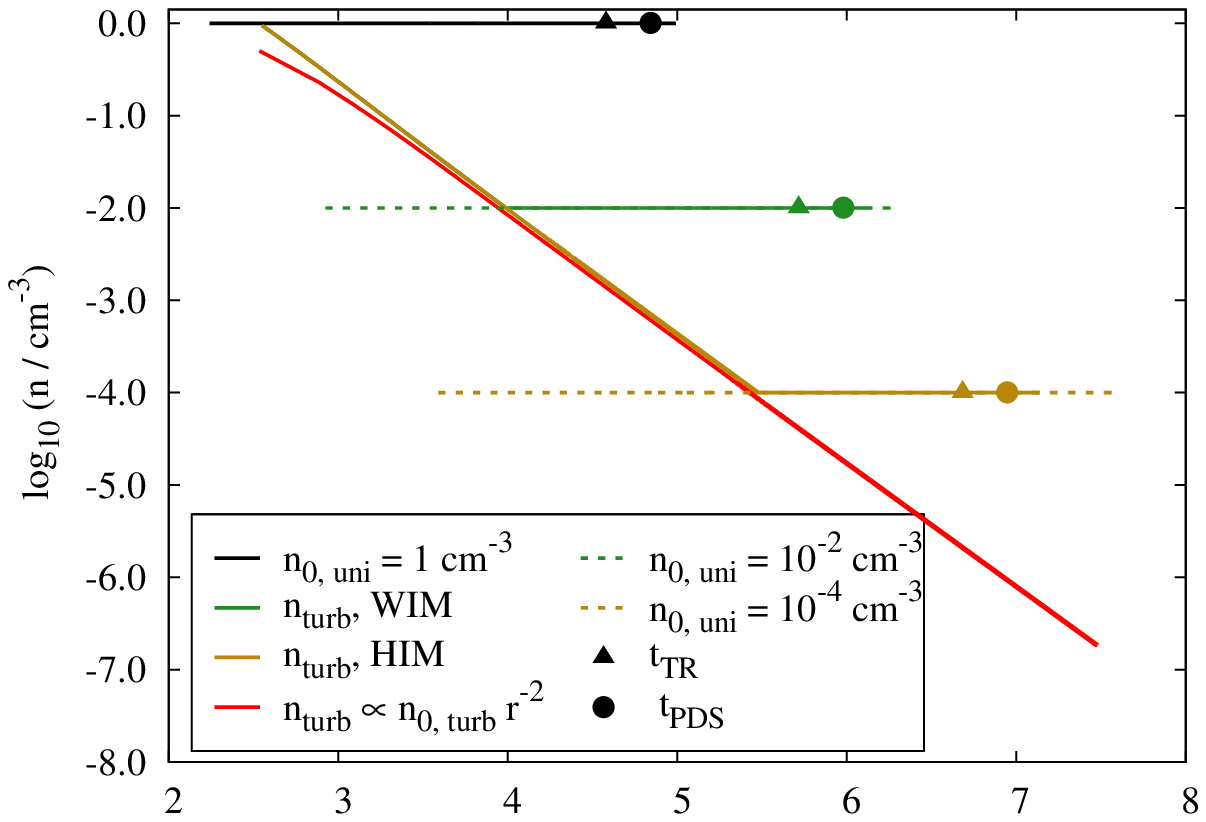}
\end{minipage} & 
\begin{minipage}[t]{0.5\textwidth}
\includegraphics[width=1.\textwidth]{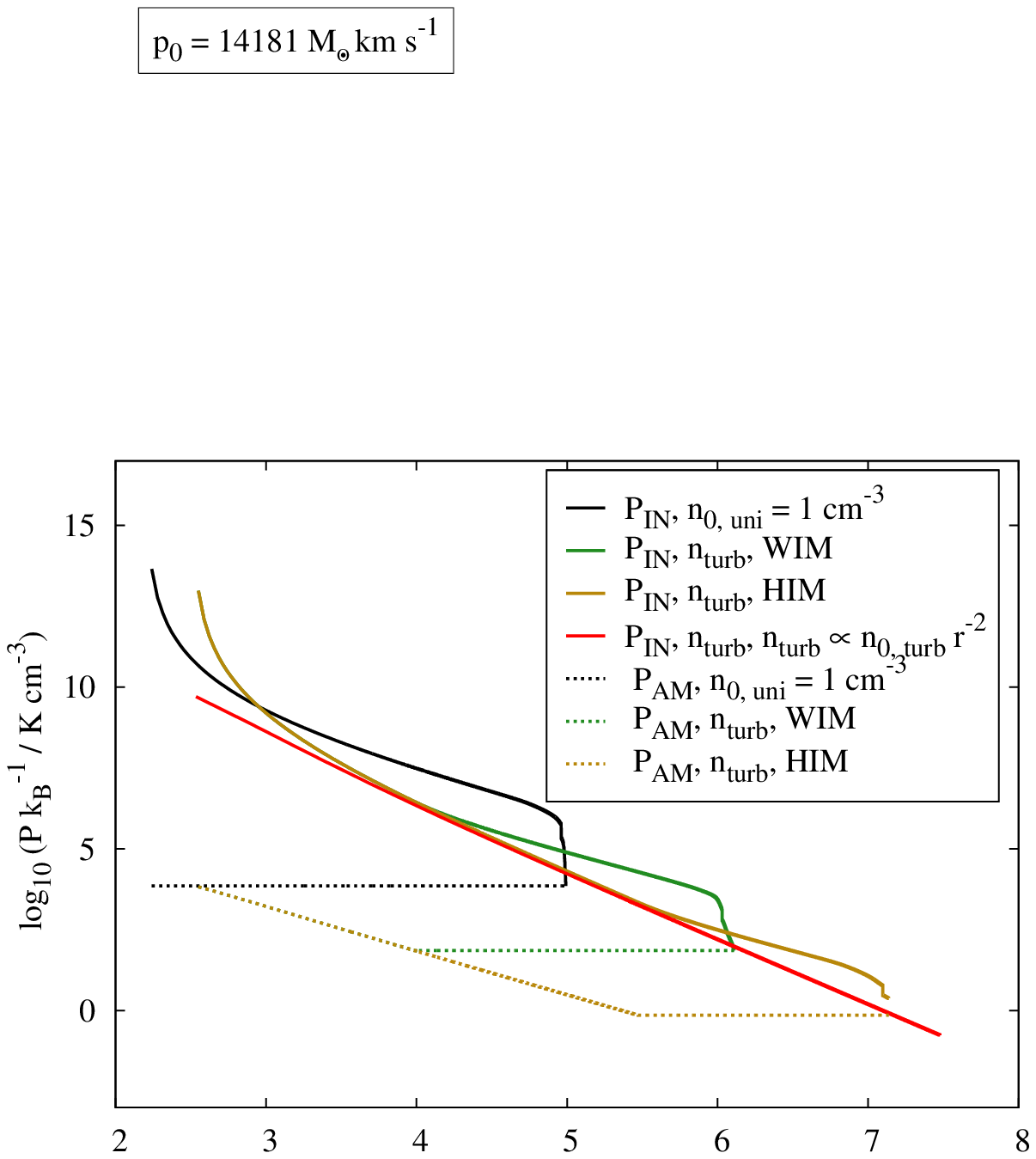}
\end{minipage} \\
\begin{minipage}[t]{0.5\textwidth}
\includegraphics[width=1.\textwidth]{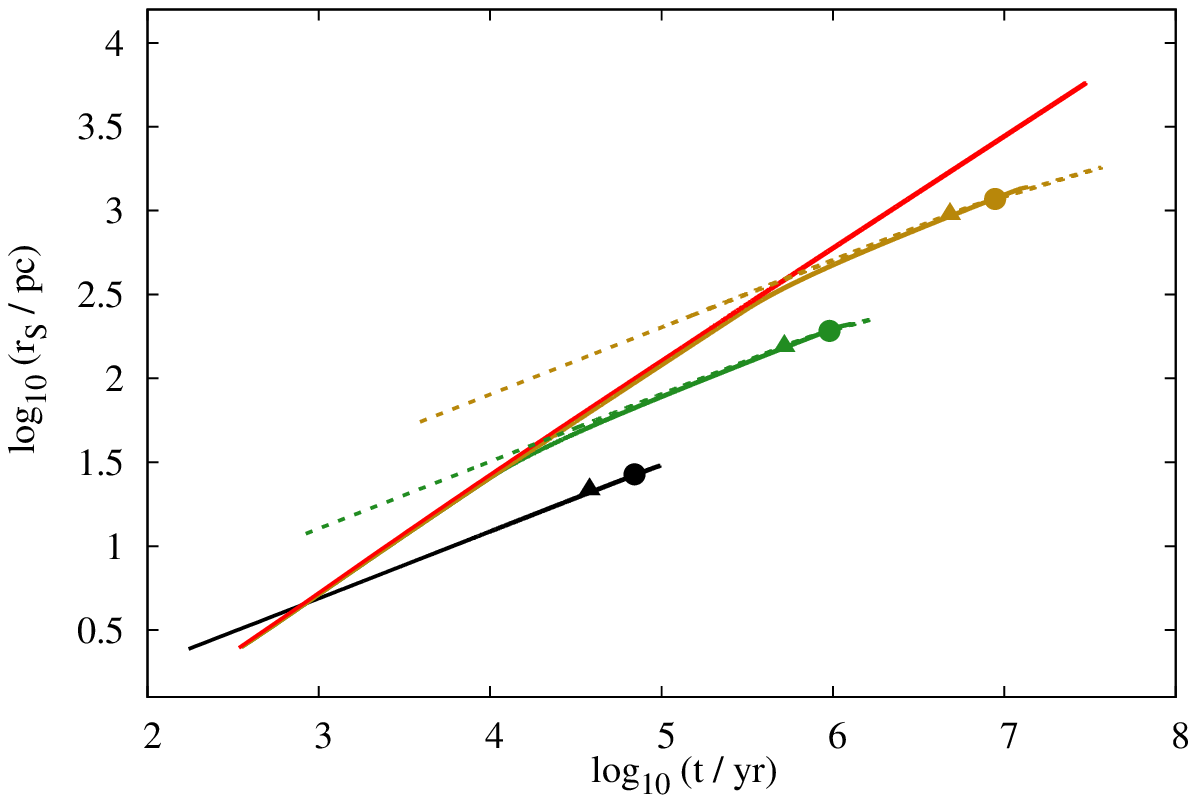}
\end{minipage} & 
\begin{minipage}[t]{0.5\textwidth}
\includegraphics[width=1.\textwidth]{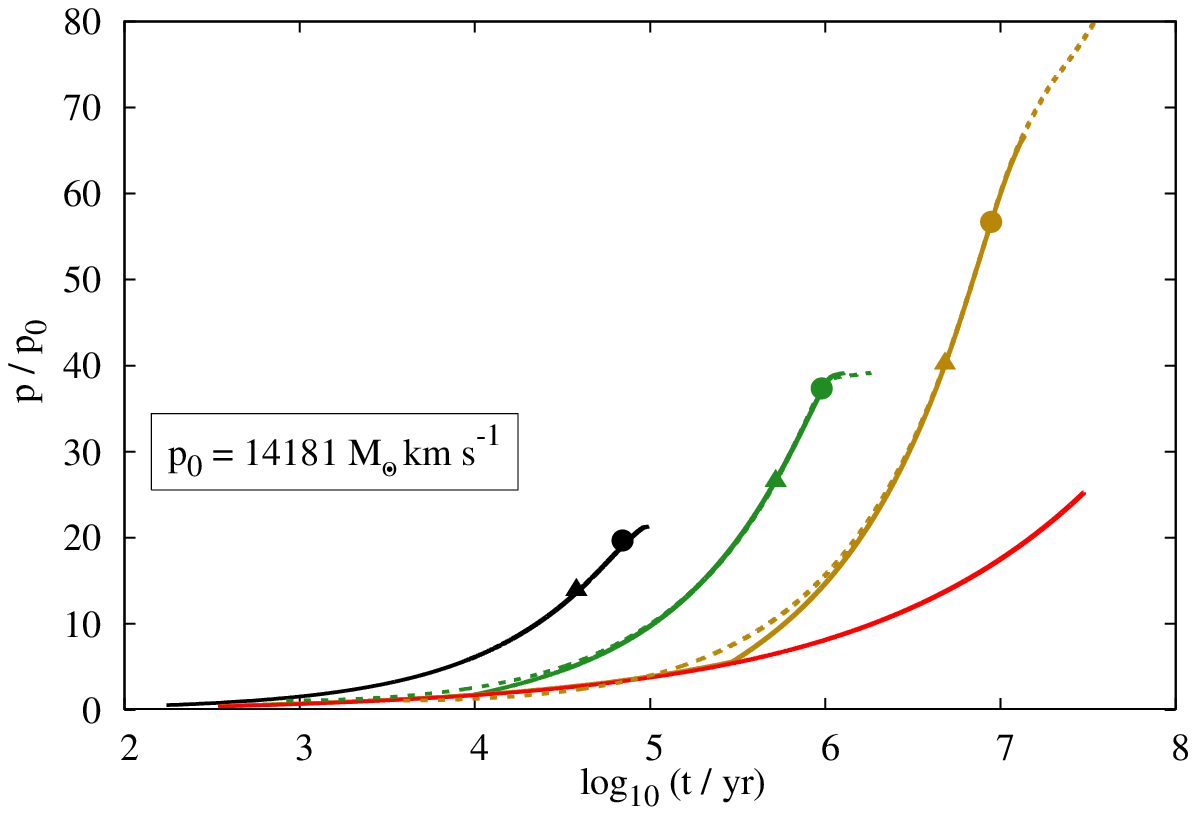}
\end{minipage} \\
\end{tabular}
\caption{Time evolution of a SNR expanding into ambient media with four different density distributions: a homogeneous (black lines) environment with a density of $n_{_{0,\rm power}}$ = 1 cm$^{3}$, media with a power-law distribution $\omega$ = 2 and density floors similar to the WIM (green lines, 7000 K, $n_{_{\rm floor}}$ = 10$^{-2}$ $\mathrm{cm^{-3}}$, $P/k_{_{\rm b}}$ = 70 $\rm cm^{-3}$ K) and the HIM (dark yellow lines, 3$\times10^{5}$ K, $n_{_{\rm floor}}$ = 10$^{-4}$ $\mathrm{cm^{-3}}$, $P/k_{_{\rm b}}$ = 30 $\rm cm^{-3}$ K) and a power-law distributed medium without a lower limit (red lines). Dashed lines correspond to homogeneous ambient media with $n_{_{0,\rm power}}$ = $n_{_{\rm floor}}$ (HIM, WIM). Triangles indicate the beginning of the TR phase, circles the onset of the PDS phase. {\it Top left panel:} Evolution of the expansion radius. {\it Top right panel:} Evolution of the internal pressures (solid lines) and the counteracting ambient pressure (dotted lines). {\it Bottom left panel:} Number density evolution at the shock front, showing the assumed density floors. {\it Bottom right panel:} Evolution of the radial momentum input. The simulation without a density floor is terminated after $\sim$ 30 Myr.} 
\label{fig:evogradover} 
\end{figure*}

In Fig. \ref{fig:evogradover} we illustrate the impact of different values of $n_{_{\rm floor}}$ ( $n_{_{\rm floor}}$ = 10$^{-2}$, 10$^{-4}$ cm$^{-3}$) on the remnant evolution in power-law environments. For comparison, we show the case of a homogeneous ambient medium with $n_{_{0,\rm power}}$ = 1 cm$^{-3}$ (black, solid line), $n_{_{0,\rm power}}$ = 10$^{-2}$ cm$^{-3}$ (green, dashed line) and $n_{_{0,\rm power}}$ = 10$^{-4}$ cm$^{-3}$ (dark yellow, dashed line). We compare the case of a SNR expanding into a warm ionized medium (WIM case; green lines) with $n_{_{\rm floor}}$ =10$^{-2}$ $\mathrm{cm^{-3}}$ ,  $T$ = 7000 K, and $P/k_{_{\rm b}}$ = 70 $\rm cm^{-3}$ K; or into a hot ionized medium (HIM case; dark yellow lines) with $n_{_{\rm floor}}$ = 10$^{-4}$ cm$^{-3}$, T = 3$\times10^{5}$ K, and $P/k_{_{\rm b}}$ = 30 $\rm cm^{-3}$ K, respectively \citep{mckee95}. A plain power-law with no density floor (red lines) is also shown. We terminate the latter simulation at 30 Myr. The density distributions are shown in the top left panel of Fig. \ref{fig:evogradover}.

In the top right panel of Fig. \ref{fig:evogradover} we show the interior pressure, $P/k_{_{\rm b}}$ (full lines) and the counteracting ambient pressure (dotted lines). Assuming an isothermal environment, the ambient pressure is directly proportional to the density distribution. The homogeneous ambient medium is isobaric, whereas in the WIM and HIM the pressure decreases with increasing radius down to the isobaric floor. The pressure in the ambient medium with a plain power-law would decrease to zero at infinity. The pressure inside the bubble decreases and drops significantly at $t_{_{\rm TR}}$ when radiation becomes important. When the ambient pressure is equal to the interior pressure, the simulation terminates at 98 kyr (homogeneous medium), 1.3 Myr (WIM) and 26 Myr (HIM).

The expansion radius of the SNR (left bottom panel) increases with lower ambient densities. In a homogeneous medium the radius is the smallest as the shock sweeps-up mass with a constant density. The power-law media with homogeneous surroundings show similar behaviour but different final radii depending on the ambient pressure. The final radius in the WIM is $\sim$ 200 pc ($t_{_{\rm MCS}}$ = 1.1 Myr) and in the HIM $\sim$ 1020 pc ($t_{_{\rm MCS}}$ = 5.6 Myr).
For the plain power-law the density drops with the radius. The counteracting swept-up mass is missing and the expansion terminates without forming a dense shell \citep{ostriker88, truelove99, petruk06}. 

The final radial momentum input (Fig. \ref{fig:evogradover}, bottom right panel) increases from 
22.9 $\rm p_{_{0}}$ in the homogeneous medium and almost doubles to 39.0 $\rm p_{_{0}}$ assuming a WIM. In the HIM the momentum input is 68.3 $\rm p_{_{0}}$. The momentum in the plain power-law environment increases continuously.

To summarize, we find that the momentum injection in a power-law environment is small compared to the uniform medium, because the decreasing density suppresses the coupling of the momentum to the gas. If the power-law environment is surrounded by a homogeneous density floor the final momentum can increase. However, the momentum input is always smaller or equal to the case of a uniform ambient medium with $n_{_{0, \rm uni}}$ = $n_{_{\rm floor}}$, independent of $n_{_{0, \rm power}}$.

\section{Blast wave evolution in wind-driven bubbles}
\label{sec_bubble}

\begin{figure*}
\begin{tabular}{|c|c|}
\begin{minipage}[t]{0.5\textwidth}
\includegraphics[width=1.\textwidth]{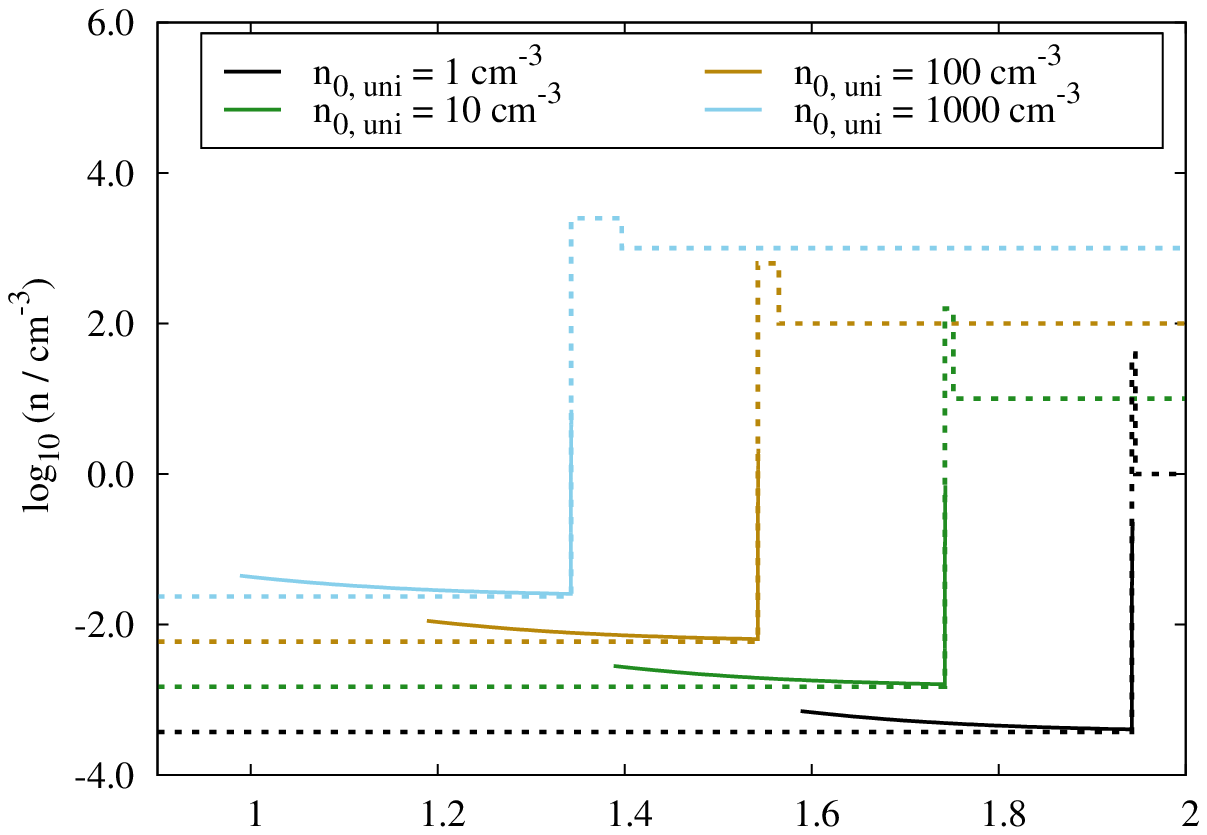}
\end{minipage} & 
\begin{minipage}[t]{0.5\textwidth}
\includegraphics[width=1.\textwidth]{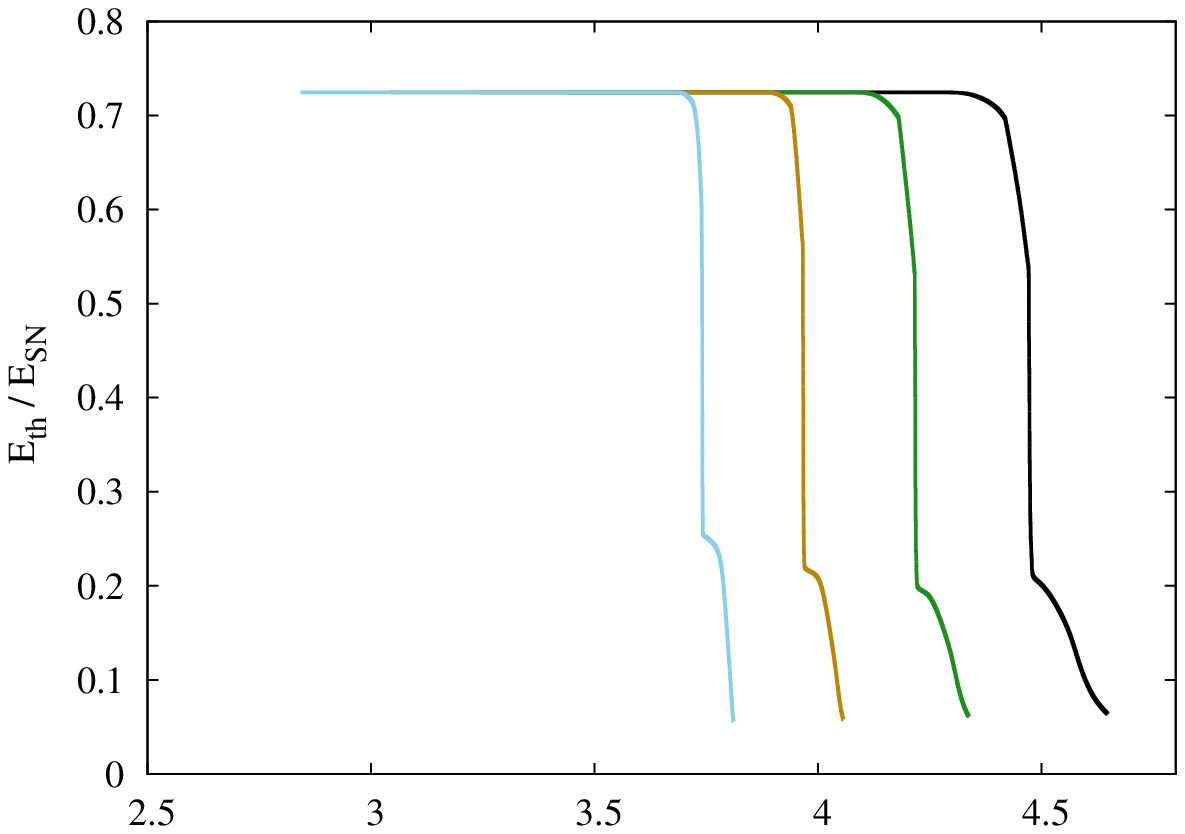}
\end{minipage} \\
\begin{minipage}[t]{0.5\textwidth}
\includegraphics[width=1.\textwidth]{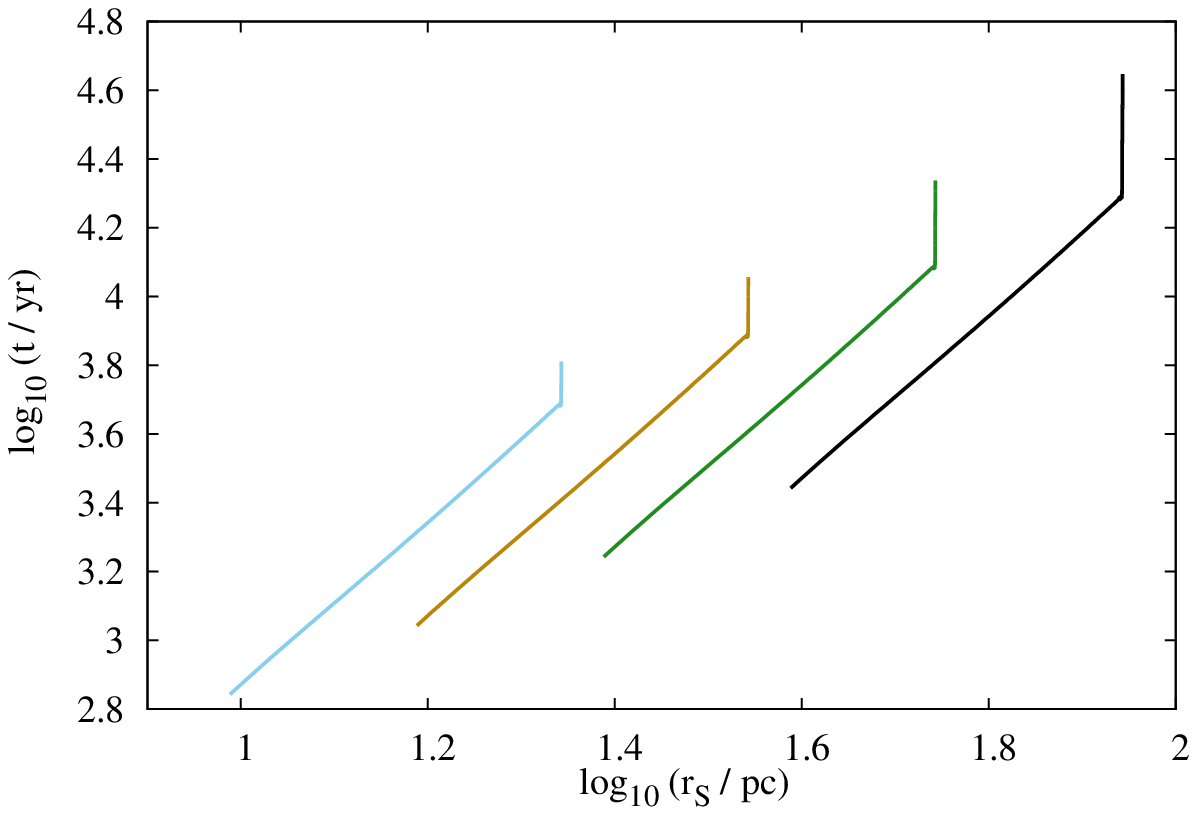}
\end{minipage} & 
\begin{minipage}[t]{0.5\textwidth}
\includegraphics[width=1.\textwidth]{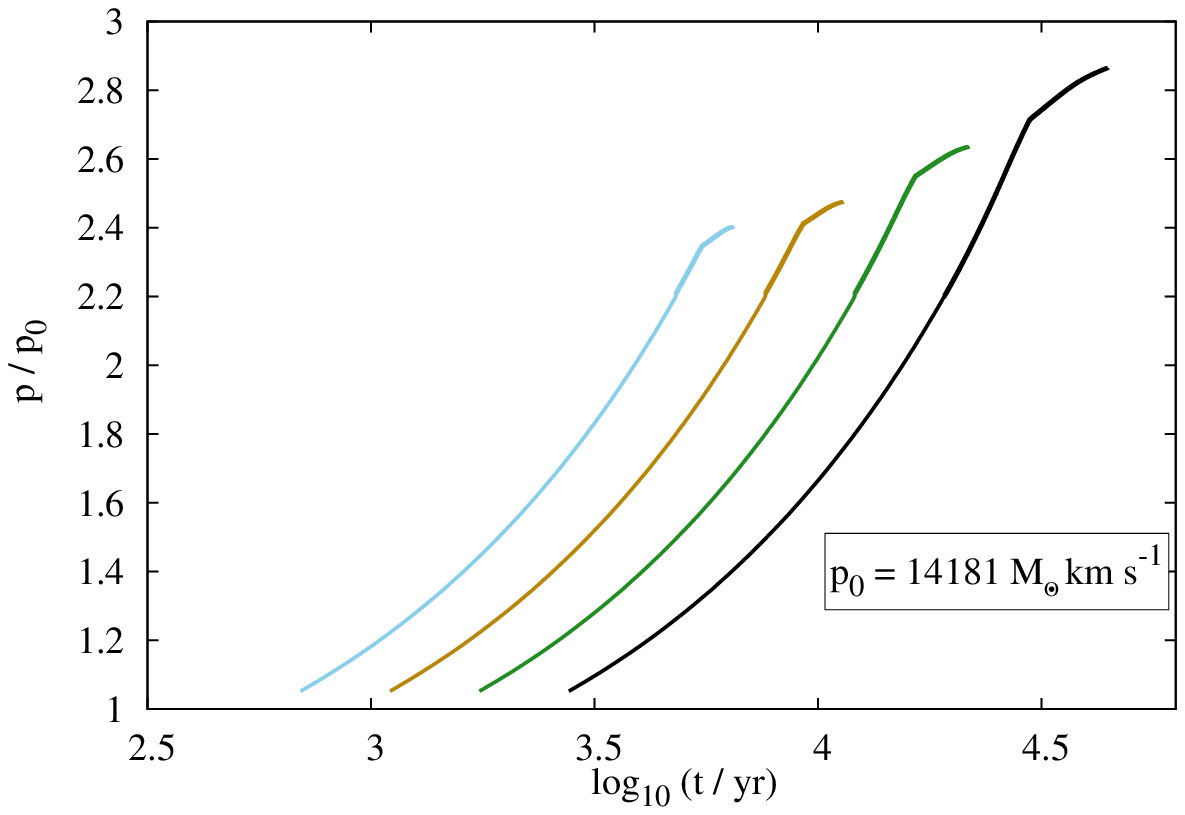}
\end{minipage} \\
\end{tabular}
\caption{Time evolution of a SNR (M$_{_{\rm eject}}$  = 2 M$_{_{\odot}}$) in a pre-existing bubble from a constant wind expanding into an initially homogeneous medium with densities of  $n_{_{0,\rm uni,}}$ = 1, 10, 100, 1000 $\rm cm^{-3}$, an initial temperature of 80 K and $P_{_{0}}/k_{_{\rm b}}$ = 80 $-$ 8 $\times 10^{4}$  $\rm cm^{-3}$ K. The density in the interior is assumed to be constant (top left panel) and in a constant density environment.  {\it Top left panel:}Radial density distribution of the pre-existing wind-blown bubble (dashed lines) and the mean density of the SNR (full lines). {\it Top right panel:} Time evolution of the normalized thermal energy. {\it Bottom left panel:} Evolution of the time over the SN shock radius. {\it Bottom right panel:} Time evolution of the radial, normalized momentum input.} 
\label{fig:evoweaver} 
\end{figure*}

\begin{figure*}
\includegraphics[width=0.8\textwidth]{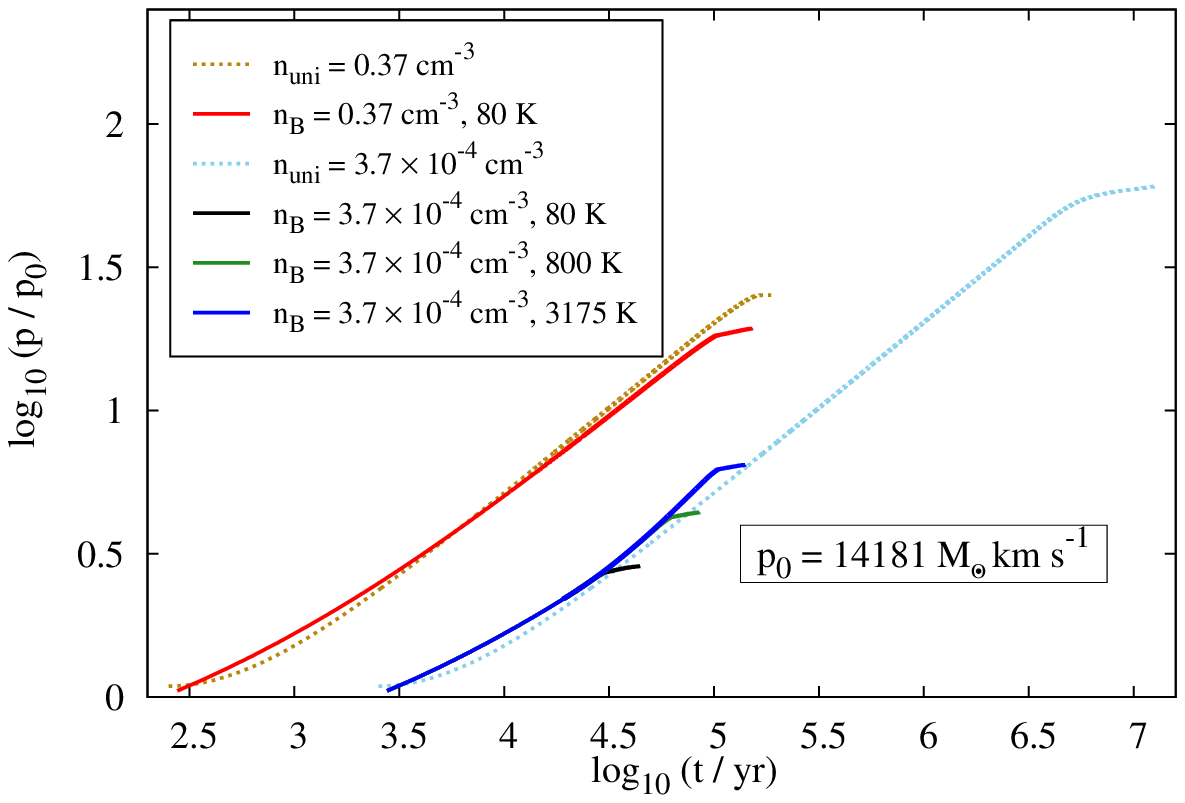}
\caption{Radial momentum of SNR in wind-blown bubbles in different initial media in comparison to uniform media (dashed lines). The densities are $n _{_{\rm B}}$ = n$_{_{\rm uni}}$ = 3.7 $\times$ 10$^{-4}$ $\rm cm^{-3}$ and 0.37 $\rm cm^{-3}$. The initial temperatures of the ambient medium, which hosts the wind-blowing star are 80, 800 and 3175 K. At the latter temperature the shock speed is equal to the sound speed of the medium.} 
\label{fig:evoweaver_temp} 
\end{figure*}

During the lifetime of a massive star strong stellar winds interact with the ambient medium and blow low-density bubbles \citep{weaver77}. The subsequent SNe explode in these bubbles and the evolution of the blast wave is modified. Here we discuss the evolution of SN blast waves in wind-blown bubbles. We assume a simple model for a constant wind expanding into an initially cold (80 K) homogeneous medium with four different initial densities ($n_{_{0,\rm uni}}$ = 1, 10, 100, 1000 $\rm cm^{-3}$). In these cold environments the wind-blown bubble expands supersonically and drives a strong shock into the ambient ISM. The shock is radiative and cools down to $T_{_{\rm s, SH }}$.

We assume a 20 M$_{_{\odot}}$ O-star with a constant wind velocity of v$_{_{\omega}}$= 2000 km s$^{-1}$ and a constant mass-loss rate of $\dot{\rm M}_{_{\omega}}$ = $10^{-7}$ M$_{_{\odot}}$ yr$^{-1}$ over a lifetime of $t_{_{B}}$ =  10 Myr. The SN has an ejecta mass M$_{_{\rm eject}}$  = 2 M$_{_{\odot}}$ \citep{puls09}. The expansion radius $r_{_{\rm s, B}}$ of a wind-blown bubble from a constant stellar wind without heat transfer is given by \citep{weaver77, pittard13}
\begin{equation}
r_{_{\rm s,B}}(t) = \left( \frac{125}{154\pi}\right)^{1/5} \left( \frac{L_{_{\rm \omega}}}{\rho_{_{0, \rm uni}}}\right)^{1/5} t^{3/5}
\end{equation}
where $ \rho_{_{0, \rm uni}}$ is the density of the initial homogeneous ambient medium with $\mu$ = 1. $L_{_{\rm \omega}}$ is the mechanical luminosity
\begin{equation}
L_{_{\rm \omega}} = \frac{1}{2}\dot{M}_{_{\rm \omega}}v_{_{\rm \omega}}^{2}.
\end{equation}

The average density $\rho_{_{\rm B}}$ within the bubble without mixing is \citep{dyson73, garciasegura95, pittard13}
 \begin{equation}
\rho_{_{\rm B}}(t) = \frac{3 \dot{M}_{_{\omega}}t}{4\pi r_{_{\rm s,B}}^{3} }.
\end{equation}

The density of the wind-shocked shell $\rho_{_{\rm s, B}}$ can be estimated by the isothermal shock jump condition ($\gamma$ = 1),
 \begin{equation}
\rho_{_{\rm s, B}} = \rho_{_{0, \rm uni}} \frac{v_{_{\rm s, B}}^{2}}{c_{_{0}}^{2}}
\end{equation}
where $c_{_{0}}$ is the sound-speed of the ambient medium with $c_{_{0}} = (\gamma P_{_{0}} / \rho_{_{0, \rm uni}})^{1/2}$. The wind bubble expands supersonically with the velocity $v_{_{\rm s, B}}$ 
 \begin{equation}
\frac{d}{dt}(r_{_{\rm s,B}}) =v_{_{\rm s, B}} =  \frac{3}{5}\frac{r_{_{\rm s,B}}}{t}.
\end{equation}

The shell thickness $\delta r_{_{\rm s, B}}$ is
 \begin{equation}
\delta r_{_{\rm s, B}} =  \frac{c_{_{0}}^{2}}{3}\frac{r_{_{\rm s, B}}}{v_{_{\rm s, B}}^{2}}.
\end{equation}

In Fig. \ref{fig:evoweaver} we show the evolution of a SN in each of the four pre-existing wind-blown bubbles. The densities inside the bubble, $n_{_{\rm B}}$, are 3.7, 14.8, 59.1 and 235.1 $\times 10^{-4}$ $ \rm cm^{-3}$ for ambient densities of $n_{_{0,\rm uni}}$ = 1, 10, 100, 1000 $\rm cm^{-3}$ (top left panel, dashed line). The interior is separated from the ambient medium by a dense shell. The density contrast of between the interior and the shell is constant with 1.5 $\times\, 10^{-5}$. The thickness of the shells are 0.7, 1.2, 1.8 and 2.9 pc. The density of the SNR follows this evolution until the evolutions stalls. 

The SN evolution in the low density interior is dominates by the ST phase, which immediately ends when the blast wave hits the dense shell (top left panel). Within $\sim$ 2 kyr 80 per cent of the initial thermal energy is radiated away, almost independently of the shell density. The remaining thermal energy is related to the hot, low-density interior of the SNR. Previous works \citep[e.g.][]{dwarkadas07} show a similar behaviour of rapid cooling at the shock boundary. Recent numerical simulations \citep{fierlinger15} point out that 1.5 per cent of the SN energy is left after the SNR stalls at the boundary. 

Initially the radial evolution (bottom left panel) is that within a homogeneous medium. For the densest ambient medium ($n_{_{0,\rm uni}}$ = 1000 $\rm cm^{-3}$) the wall of the wind-blown cavity is reached after $\sim$ 4.9 kyr and 22.0 pc, while it takes $\sim$ 12.2 kyr and 87.6 pc for $n_{_{0,\rm uni}}$ = 10 $\rm cm^{-3}$. The final radius corresponds to the inner radius of the bubble.

The density distribution of the wind-bubble is assumed to be static and the shell has no momentum. While in the ST phase, the momentum input by the SN is small because of the low gas density within the bubble. Once the remnant reaches the shell, which is massive compared to the swept-up mass from the SN, it cools quickly and cannot accelerate the shell. As a result the evolution of the SNR stalls. The final momentum input (bottom right panel) lies between $\sim$ 2.4 and 2.9 $\rm p_{_{0}}$.

The density difference between the interior and the shock as well as the density of the wind-blown shell itself determine the final radial momentum. Assuming isothermal behaviour, the ambient temperature of the initial environment is linked to the shell temperature, which again effects the thickness of the shell. Therefore, in Fig. \ref{fig:evoweaver_temp} we show the influence of densities, $n _{_{\rm B}}$, and the temperature of the ambient interstellar medium on the momentum input. We choose $n _{_{\rm B}}$ = 3.7 $\times$ 10$^{-4}$ and 0.37 $\rm cm^{-3}$, where the first corresponds to an wind-blown bubble with an initial density $n _{_{0, \rm uni}}$ = 1 $\rm cm^{-3}$ and the latter corresponds to a bubble which is filled by ionised gas as would be the case for an HII region. We increase the temperatures from 80 K to 800 K and to the temperature (3175 K), which corresponds to $v_{_{\rm s, B}}$ = $c_{_{0}}$. The dashed lines show the momenta of SNe in uniform media with n$_{_{\rm B}}$ = $n _{_{0, \rm uni}}$.

For the low density case ($n _{_{\rm B}}$ = 3.7 $\times$ 10$^{-4}$ $\rm cm^{-3}$) we show how the final momentum increases with temperature from 2.9 $\rm p_{_{0}}$ at 80 K to 4.4 $\rm p_{_{0}}$ at 800 K and up to 6.5 $\rm p_{_{0}}$ at 3175 K. At a higher interior density ($n _{_{\rm B}}$ = 0.37 $\rm cm^{-3}$) the momentum in the cold (80 K) ambient medium is 19.3  $\rm p_{_{0}}$ and is comparable to the corresponding homogeneous medium. Recent numerical results of SNe exploding into bubbles blown by a stellar wind and ionizing radiation give a factor of $\sim$ 10 \citep{geen15}.

This shows that the ambient density and temperature are essential for the evolution of a SNR in a wind-blown bubble. Higher temperatures broaden the wind-blown shell and reduce the density contrast. This results in a lower cooling and an increase of radial momentum \citep[e.g.][]{walch14a}. The influence of the wind-blown bubble on the evolution of the SNR diminishes as the swept-up mass increases compared to the mass of the shell. SNR with a high density inside the bubble and a small difference between the swept-up mass and the mass of the wind-blown shell show a behaviour that is comparable to a uniform medium with that bubble density.

\section{Blast wave evolution in turbulent environments}
\label{section5}
We study the evolution of a SNR expanding in a more realistic ambient medium, which is subject to isothermal, supersonic turbulence \citep{klessen98, klessen00, kainulainen09, schneider11, federrath13}. Numerical simulations suggest that the volume-weighted density PDF of gas shaped by isothermal turbulent motions can be described by a log-normal distribution \citep{vazquezsemadeni93,  padoan97, nordlund97, federrath08} ,
\begin{equation}
\label{eq:lognormal}
q(z)=\frac{1}{\sqrt{2\pi \sigma_{\ln\rho}^{2}}} \exp \left[-\frac{(z-\bar{z})^ {2}}{2\sigma_{\ln\rho}^{2}}\right],
\end{equation}
where $z=\ln(\rho/\rho_{_{0,\rm turb}})$ with a mean density of the gas $\rho_{_{0,\rm turb}}$. The median is $\bar{z}= - \sigma^{2}_{_{\ln\rho}}/2$ \citep{vazquezsemadeni94, thompson14}. The dispersion of the density distribution $\sigma_{\ln\rho}^{2}$ can be related to the Mach number $\mathcal{M}$ of turbulent motions \citep{federrath08, thompson14},
\begin{equation}
\label{eq:sigma}
\sigma^{2}_{_{\ln\rho}} \sim \ln(1+b^{2}\mathcal{M}^{2}).
\end{equation}
The turbulent driving factor $b$ is assumed to be 0.5 with a thermal mix of divergence free (solenoidal) and curl free (compressive) turbulence \citep{federrath08, brunt10, krumholz14}.

The volume density PDF can also be related to the surface density PDF $\sigma_{_{\ln\Sigma}}$ \citep{brunt10, brunt10b, brunt10c}. In this case, the dispersion reads
\begin{equation}
\sigma^{2}_{_{\ln\Sigma}}=\ln(1+Qb^{2}\mathcal{M}^{2}).
\end{equation}
with the conversion factor
\begin{equation}
Q=\sigma_{\ln\Sigma}^{2}/\sigma_{\ln\rho}^{2}.
\end{equation}

\subsection{Approximating the turbulent structures of the ambient medium}
\label{section5_1}

We adopt our model to compute the SNR evolution in turbulent ambient media, where the density structure is described by the log-normal PDF in Eq. \ref{eq:lognormal}. Since the blast wave evolution is primarily determined by the mean density of the swept-up material \citep{ostriker88, padoan97}, we assume that small-scale density fluctuations along the radial direction of the SNR have a negligible effect on the evolution. We assume that in different directions, the blast wave will encounter gas with different mean densities. 

In this simplified model we abstain from following winding shock fronts between structures with a large density gradient \citep[e.g.][]{martizzi14} or interaction between different radial directions. The first constraint arises from the simple set of equations used in our model. It is not designed to follow the dynamical evolution but gives a statistical expectation of SNR in turbulent media. For the latter we assume no physical interactions between the different cones and assume that during the ST and TR phase the radially outwards directed velocities of the SNR are large and the interaction has minor effects. At later phases the extent of the different radial directions is sufficient to neglect an interacting boundary. 

\begin{figure}
\includegraphics[width=0.45\textwidth]{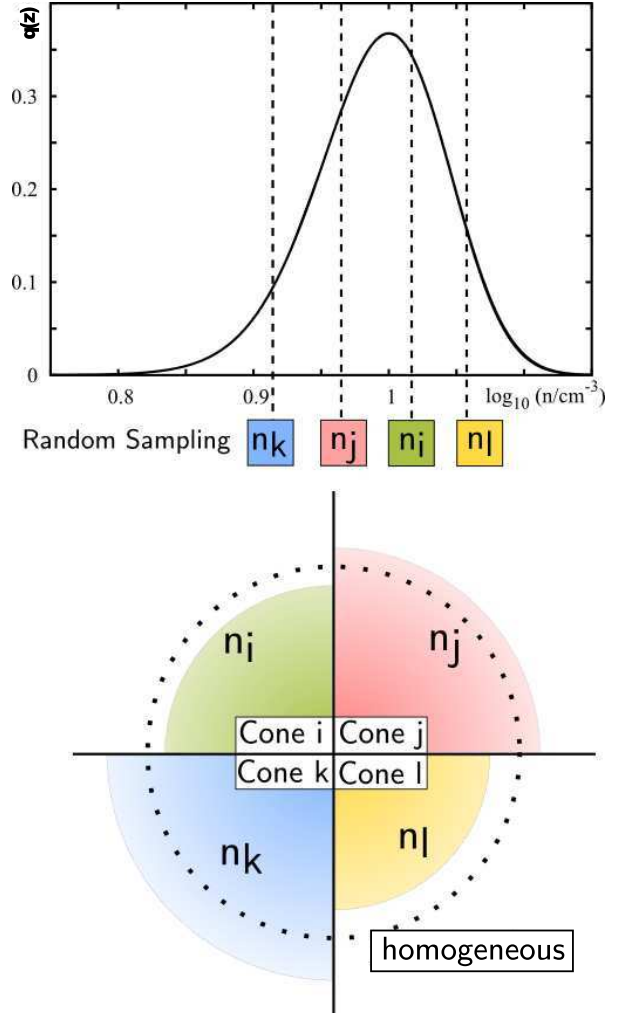}
\caption{Schematic representation of the model for the blast wave evolution into a turbulent medium. \textit{Top panel}: Sampling of densities from a log-normal PDF, which represents turbulent density structures. The number of sampling points corresponds to the number of cones with equal- surface areas.  \textit{Bottom panel}: Homogeneously assigning the densities to the cones.  The blast wave evolution is then completed for each cone separately. The total momentum input is the sum of the individual solutions. } 
\label{fig:explanation}
\end{figure}

To model the mean densities in different radial directions, the ambient medium in our model is discretized (see Fig. \ref{fig:explanation}, bottom panel) into $N_{_{\rm cones}}$ cones. The cones are defined by equal solid angles and have equal surface areas and volumes. For each cone, we randomly draw a mean density, $n_{_{i}}$, from the log-normal density distribution and run the 1-dimensional model of the evolution of a SNR for an uniform medium (see  Section \ref{section2}). The total momentum $p_{_{\rm turb}}$  injected by a SN in this pseudo 3-dimensional turbulent medium is derived from the sum over all cone momenta $p_{i}$, 
\begin{equation}
\sum_{i}^{N_{_{\rm cones}}}p_{_{\rm i}} = p_{_{\rm turb}}. 
\end{equation}
Each cone is initialised with the same fraction of the total SN energy, i.e. $E_{_{\rm SN}} / N_{_{ \rm cones}}$. As the expansion radius in each cone is different, the symmetry of the SN bubble is broken \citep{walch14a}.
\\

\begin{figure}
\includegraphics[width=0.50\textwidth]{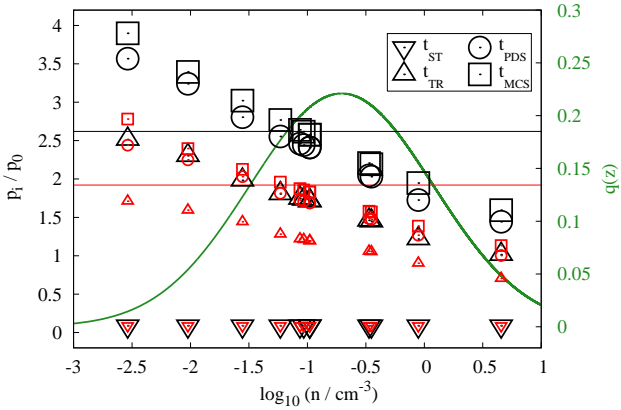}
\caption{Example for the SN momentum injection in a turbulent medium sampled with 12 cones. The number densities are randomly drawn from a log-normal PDF with a mean number density $n_{_{0,\rm turb}}$ = 1 $\mathrm{cm^{-3}}$ and a turbulent Mach number $\mathcal{M}$ = 10. We show the values at $t_{_{\mathrm{ST}}}$ (upside down triangles), $t_{_{\mathrm{TR}}}$ (triangles), $t_{_{\mathrm{PDS}}}$ (circles) and $t_{_{\mathrm{MCS}}}$ (squares) within a ionized ambient medium ($\mu_{_{\rm i}}$, red symbols) and an atomic ($\mu_{_{\rm a}}$, black symbols). The individual radial momentum for each cone $p_{_{\rm i}}$ is shown as a function of the sampled density $n$. At $t_{_{\mathrm{PDS}}}$ the mean momentum per cone is 2.6 [1.9] $\rm{p_{_{0}}}$ (black [red] horizontal line). The underlying log-normal PDF is indicated with a green line. }
\label{fig:mv3}
\end{figure}

In Fig. \ref{fig:mv3} we show results using 12 cones, which is the minimum number needed to divide the unit sphere into equal surface area pixels \citep[see ][]{gorski05}. With $N_{_{\rm cones}}$ = 12 the log-normal PDF is not well sampled (see Section \ref{section_acc} for a further discussion. The turbulent Mach number is 10 and the mean number density of the ambient medium, is $n_{_{0,\rm turb}}$ = 1 $\mathrm{cm^{-3}}$. The sampled densities $n_{_{i}}$ have values between $\mathrm{3 \times 10^{-3}\,cm^{-3}}$ and $\mathrm{4.5\,cm^{-3}}$ according to a PDF with a width of $\sigma_{_{\ln\rho}}$ = 1.8 for $\mathcal{M}$ = 10. Fig. \ref{fig:mv3} shows the equal initial momenta (upside down triangles) as well as the individual momenta $p_{_{\rm i}}$ at the end of the individual ST (triangles), TR (circles) and PDS (squares) phase for a neutral ($\mu_{_{\rm a}}$, black) and ionized ($\mu_{_{\rm i}}$, red) medium for all mean cone densities $n_{_{\rm i}}$ (green line and corresponding y-axis on the right-hand side).

The mean momentum per cone, $\left< p_{_{\rm i}} \right>$,
\begin{equation}
 \left< p_{_{\rm i}} \right> = \frac{p_{_{\rm turb}}}{N_{_{\rm cones}}},
\end{equation}
in a neutral [ionized] medium at $t_{_{\rm TR}}$ is 1.7 [1.2] $\rm{p_{_{0}}}$, which increases up to 2.4 [1.7] $\rm{p_{_{0}}}$ at $t_{_{\rm PDS}}$ ($\rm{p_{_{0}}}$ = 14181 $\rm M_{\odot}\,km\,s^{-1}$). At $t_{_{\mathrm{MCS}}}$ the mean momentum per cone is 2.6 [1.9] $\rm{p_{_{0}}}$, as indicated by the black horizontal line (red line for ionized ambient medium). This corresponds to a total momentum of 31.2 [22.8] $\rm{p_{_{0}}}$ ($\rm 2.16 \times  10^{5}\,M_{\odot}\,km\,s^{-1}$). Note that $t_{_{\rm TR}}$, $t_{_{\rm PDS}}$ and $t_{_{\rm MCS}}$ are different for cones with different densities. However, since the momentum stays constant after $t_{_{\rm MCS}}$, $p(t_{_{\rm MCS}})$ is considered as the final momentum.

The blast wave simulation in a homogeneous medium with $n_{_{0,\rm uni}}$ = 1 $\mathrm{cm^{-3}}$ injects 22.3 [16.4] $\rm{p_{_{0}}}$ of momentum. Therefore, the increase in momentum is a direct consequence of turbulence. For higher $\mathcal{M}$, the PDF becomes broader. The blast wave encounters  more low density regions, which are subject to less radiative cooling and allow for a higher momentum injection.

\subsection{Accuracy of the model}
\label{section_acc}
\begin{figure}
\includegraphics[width=0.49\textwidth]{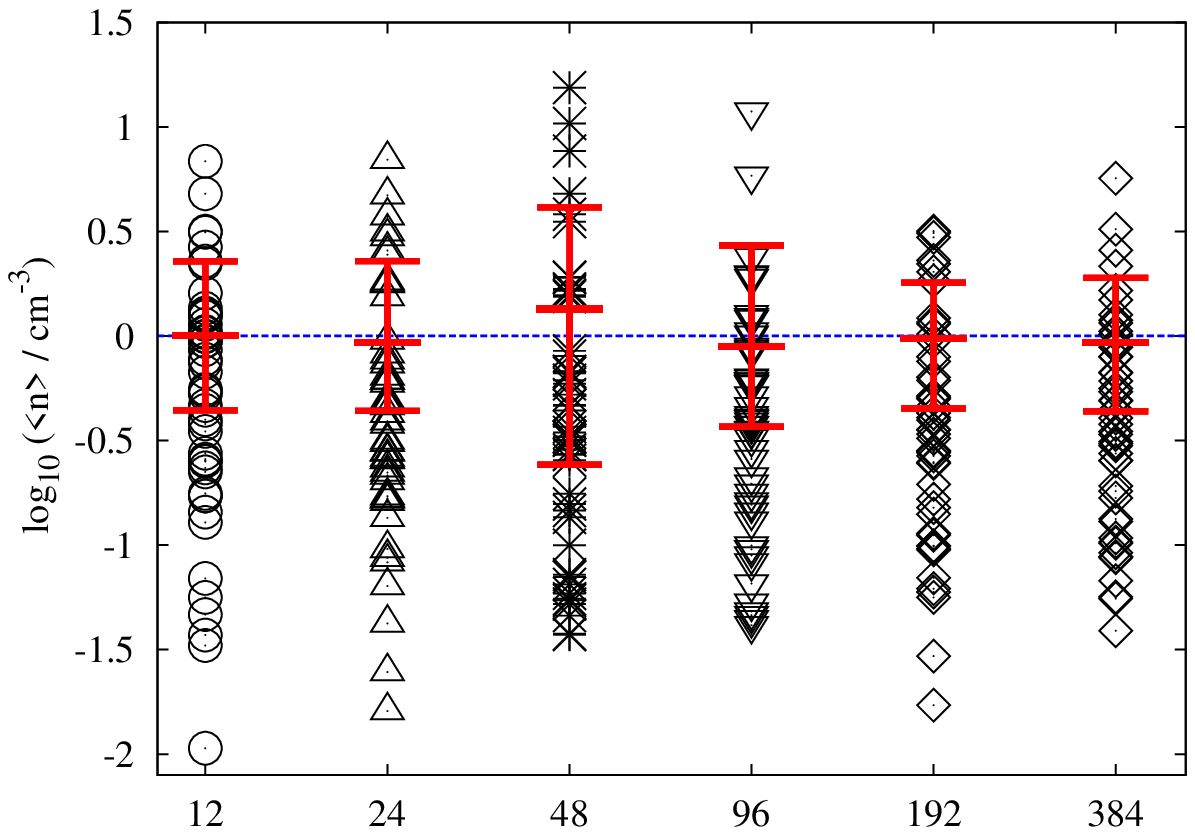}
\includegraphics[width=0.49\textwidth]{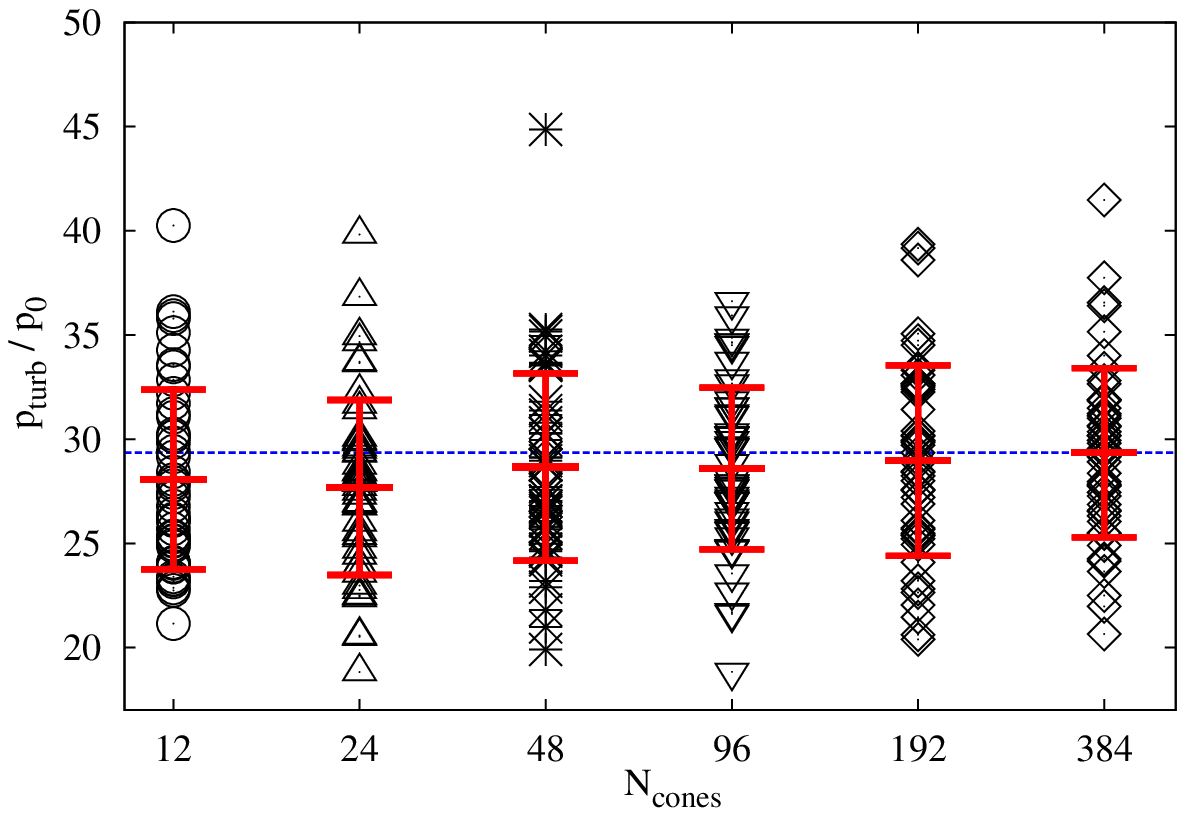}
\caption{Effect of the number of cones $N_{_{\mathrm{cones}}}$ on the accuracy of the turbulent SN model for the mean density (top panel) and momentum input (bottom panel). The number densities are randomly sampled from a log-normal PDF with a fixed mean density $n_{_{0,\rm turb}}$ = 1 $\mathrm{cm^{-3}}$ and Mach number $\mathcal{M}$ = 10. Each of the 6 data sets consists of 50 SN simulations. Mean values and the standard deviation are shown in red. The mean ambient density (blue line; top panel) is well sampled and the momentum injection converges to 29.4 $\rm{p_{_{0}}}$ (blue line; bottom panel).}
\label{fig:healpix}
\end{figure}

The fidelity of the SN model depends on the number of sampled densities, i.e. $N_{_{\rm cones}}$. We need a sufficient number in order to accurately represent the underlying density distribution.

We compute the evolution of 50 individual SN explosions in turbulent media, each with an increasing number of equal-volume cones (sampling points of the PDF) from 12 to 384. For each of the 50 runs we use a different random seed to sample the number densities in each cone from the log-normal density PDF with $n_{_{0,\rm turb}}$ = 1 $\mathrm{cm^{-3}}$ and $\mathcal{M}$ = 10.

Fig. \ref{fig:healpix} presents all 6 sets ($N_{_{\rm cones}}$ = 12, 24, 48, 96, 192, 384; different symbols) with 50 SN simulations each. In the top panel the sampled mean densities of the individual simulations, $\left< n \right> = \sum_{\rm i}^{N_{\rm cones}} n_{_{\rm i}} / N_{_{\rm cones}}$ are shown. Independent of the numbers of cones the mean ambient density ($n_{_{0,\rm turb}}$ = 1 $\mathrm{cm^{-3}}$; blue dashed line) is well sampled by the overall mean of the individual simulations (red bars). The variance decreases from 1.2 to 0.9 with increasing number of cones from 12 to 384. 

The bottom panel shows the final momentum $p_{_{\rm turb}}$ (normalized to the initial momentum) of the same simulations. The overall mean converges to 29.4 $\rm{p_{_{0}}}$ at the highest numbers of cones (blue dashed line). The variance is similar in all runs at about 4 $\rm{p_{_{0}}}$. 

To summarize, we show that the combination of high-$\mathcal{M}$-turbulence and small $N_{_{\mathrm{cones}}}$ may not accurately represent the turbulent PDF structure. Individual realizations might over/under-predict the mean densities but larger samples and a higher number of cones reduced the variance in the mean density and the momentum input.

\subsection{Momentum distribution in turbulent media}
\label{section5_2}

\begin{figure*}
\includegraphics[width=0.80\textwidth]{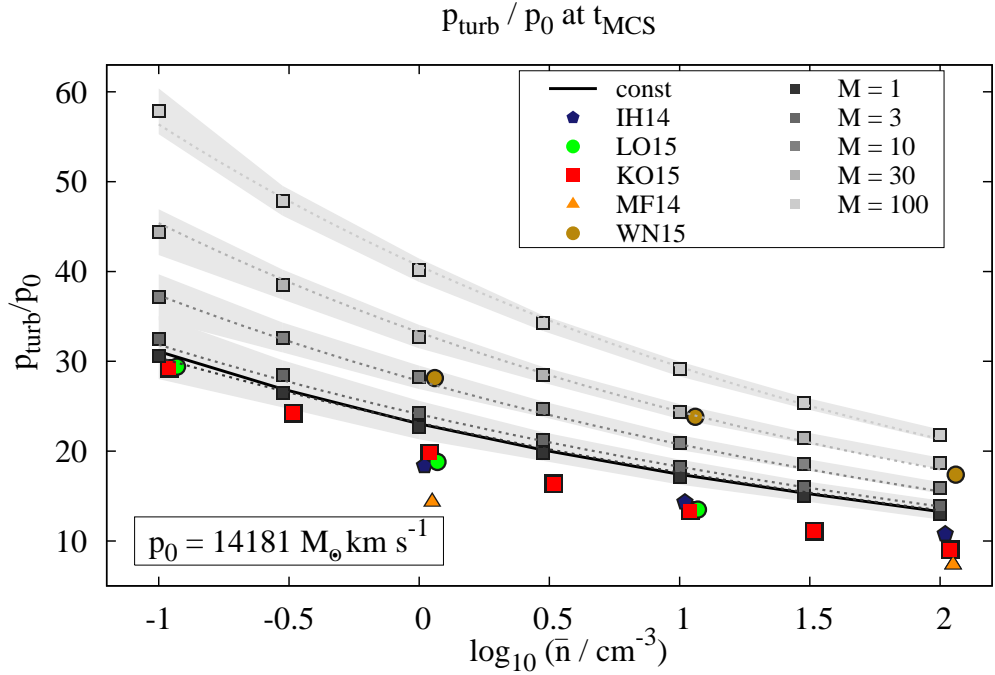}
\caption{Final (at $t_{_{\mathrm{MCS}}}$) radial momentum input $p_{_{\rm turb}}$ to turbulent media with mean densities in the range of $n_{_{0,\rm turb}}$ = 0.1 $-$ 100 $\mathrm{cm^{-3}}$ and Mach numbers increasing from $\mathcal{M}$ = 1 $-$ 100 (different grey squares). The environment of each SN is separated into 195 cones.  Each data point corresponds to the mean of 20 realizations and a standard deviation (grey shaded areas). We add recent numerical simulations from \citet[][here IH14, blue pentagons]{iffrig14}, \citet[][here KO15, red squares]{kim14}, \citet[][here MF14, orange triangles]{martizzi14} and \citet[][here WN15, dark yellow circles]{walch14a} and \citet[][here LO15, green circles]{li15}. For better visibility these symbols are shifted to the right of the corresponding number density.}
\label{fig:momdens}
\end{figure*}

We perform simulations of SNRs in turbulent media with mean densities of $n_{_{0,\rm turb}}$ = 0.1 $-$ 100 $\mathrm{cm^{-3}}$, and Mach numbers, $\mathcal{M}$ = 1 $-$ 100. Based on the previous section, we decided to use sets of 20 realizations for each turbulent setup with $N_{_{\mathrm{cones}}}$ = 192 and evaluate the total radial momenta up to $t_{_{\mathrm{MCS}}}$ of the cone with the lowest density cone (Fig. \ref{fig:momdens}).

The mean shell momenta lie between 13.0 $\rm{p_{_{0}}}$ ($n_{_{0,\rm turb}}$ = 100 $\mathrm{cm^{-3}}$, $\mathcal{M}$ = 1) and 30.6 $\rm{p_{_{0}}}$ ($n_{_{0,\rm turb}}$ = 0.1 $\mathrm{cm^{-3}}$, $\mathcal{M}$ = 1). Higher supersonic turbulence ($\mathcal{M}$ = 100) boosts the momentum by 60 per cent ($n_{_{0,\rm turb}}$ = 100  $\mathrm{cm^{-3}}$) up to 88 per cent ($n_{_{0,\rm turb}}$ = 0.1  $\mathrm{cm^{-3}}$) compared to the low-$\mathcal{M}$-turbulence value.

The radial momentum input of a single SN in a turbulent medium can be quantified in terms of the mean density and the width (Mach number) of the underlying density PDF: 
\begin{multline}
p_{_{\rm turb}}/\mathrm{p_{_{0}}}\ =23.07\, (n_{_{0,\rm turb}}/ 1\,\mathrm{cm^{-3}})^{-0.12}\\ + 0.82 (\ln(1+b^{2}\mathcal{M}^{2}))^{1.49} (n_{_{0,\rm turb}}/ 1\ \mathrm{cm^{-3}})^{-0.17}.
\end{multline}
The first term corresponds to the momentum transfer from a single SN into a homogeneous medium. The second term depends on a combination of the turbulent Mach number (width of the PDF) and the mean density. The factor in the first term is higher compared to the value (22.44) obtained for the uniform medium. The difference results from the additional turbulent term. The fit was generated over all data points by a Bees algorithm coupled with Levenberg-Marquardt provided by the fitting tool MAGIX \citep[$\chi^{2} \sim 8; $][]{bernst11, moeller13}.


In Fig. \ref{fig:momdens} we compare our results to direct, 3-dimensional (magneto-) hydrodynamical simulations from different authors, namely, \citet{iffrig14}, , \citet{martizzi14},\citet{kim14}, \citet{li15} and \citet{walch14a} (coloured symbols). We compare at times similar to our $t_{_{\rm MCS}}$. As the methodology for setting up the ISM conditions varies from author to author, we explain each set of simulations in more detail. 

\citet[][dark blue diamonds]{iffrig14} simulate SNR in highly-resolved (maximum grid resolution 0.05 pc)  turbulent MCs with magnetic fields, self-gravity and a cooling function similar to Eq. \eqref{eq:cooling}. The initial conditions for the SN explosion evolve from a spherical cloud with a density gradient $\propto$ r$^{-2}$ embedded in a low density environment. The assumed velocity field in the MC represents a Kolmogorov spectrum with a random component. The authors conclude that the influence of magnetic fields is small, rather the position and, therefore, the ambient density of the SN in the MC is determining the final momentum. It is well approximated by the solution of 3-dim SNR simulations in homogeneous medium with 18 $\rm{p_{_{0}}}$ for $n_{_{0}}$ = 1 $\mathrm{cm^{-3}}$. 

\citet[][red squares]{kim14} pre-evolve the ambient medium from a thermally unstable state with small density perturbations. The SN explodes into a two-phase environment in pressure balance. The fitted final momentum input is comparable to SNe in homogeneous media. The difference to our final momentum in low-$\mathcal{M}$-turbulent environments is smaller than 15 per cent.

\citet[][dark yellow circles]{walch14a} use a SPH particle code to perform highly-resolved (maximum resolution 0.1 M$_{_{\odot}}$) hydrodynamic simulations with interpolating cooling tables by \citet[][ for $T$ $\ge$ 10$^{4}$ K]{plewa95} and the cooling function from \citet[][ for $T$ $<$ 10$^{4}$ K]{koyama02}. The ambient medium is initialized with fractal sub-structures, which represent a log-normal density pdf. The resulting variance is translated to a turbulent Mach number, $\mathcal{M}$ = 4.4 \citep{walch11b}. The normalized final momentum $p$ = 25.6 $\rm{p_{_{0}}}$ is $\sim$ 9 per cent higher compared to values obtained from our SN model ($n_{_{0,\rm turb}}$ = 1 $\mathrm{cm^{-3}}$, $\mathcal{M}$ = 4.4.

\citet[][orange triangles]{martizzi14} perform hydrodynamic simulations in an ambient medium with a log-normal density field but only cooling by \citet{sutherland93} at temperatures above 10$^{4}$ K. The variance of the distribution uses a parametrization by \citet{lemaster09}. The spatial correlations are parametrized by a Burgers power-spectrum. The initial velocity field is set to zero. Within these structures (maximum grid resolution 0.05 pc) the SNR evolves along the path of least resistance but cools significantly (down to 10$^{4}$ K) when dense structures are hit and merge with the shock. This results in a final momentum input of of 7.3 $\rm{p_{_{0}}}$ in a supersonic environment ($\mathcal{M}$ = 30, $n_{_{0,\rm turb}}$ = 100  $\mathrm{cm^{-3}}$), which is lower than the performed fiducial simulation in a homogeneous medium. The final value is $\sim$ 2.6 lower than a similar simulation with our model.

\citet[][green circles]{li15} creates an (artificial) environment with randomly distributed cold clouds and hot inter-cloud medium with a SN in the centre. The results show no distinctive phases and an expansion between the cold and dense regions on a path of least resistance. Initially the radial momentum input is lower, than the homogeneous comparison and shows an increasing power-law behaviour with radius. As the shock expands further it interacts with the medium in non-radial directions. At the end the momentum is almost constant and similar to values from uniform media.
 The momentum of the homogeneous runs (18.8 $\rm{p_{_{0}}}$) compares with the input from structured media at later phases of 17.7 $\rm{p_{_{0}}}$ ($n_{_{0,\rm turb}}$ = 1 $\mathrm{cm^{-3}}$).

To summarize, we find that momentum input from low-$\mathcal{M}$-turbulent structures is comparable to SNR in homogeneous media. We find similar values compared to different 3-dimensional numerical simulations, under the assumption of an atomic medium. We show that high-$\mathcal{M}$-turbulent structures boost the radial momentum input. We conclude that turbulence could be important for the momentum input. However, more 3-dimensional models with very high resolution will be required to address the impact of a highly turbulent substructure.

\subsection{Velocity-mass distribution in turbulent media}
\label{section5_3}
\begin{figure}
\includegraphics[width=0.49\textwidth]{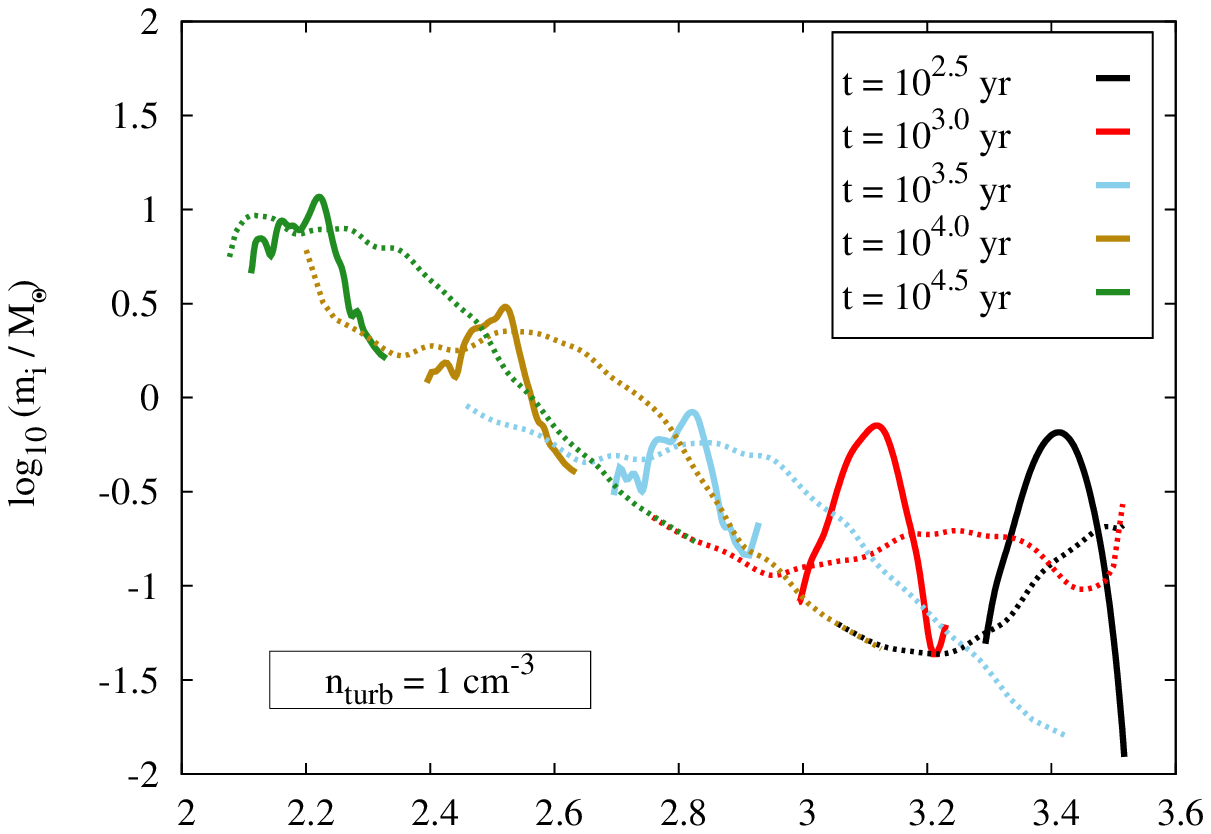}
\includegraphics[width=0.49\textwidth]{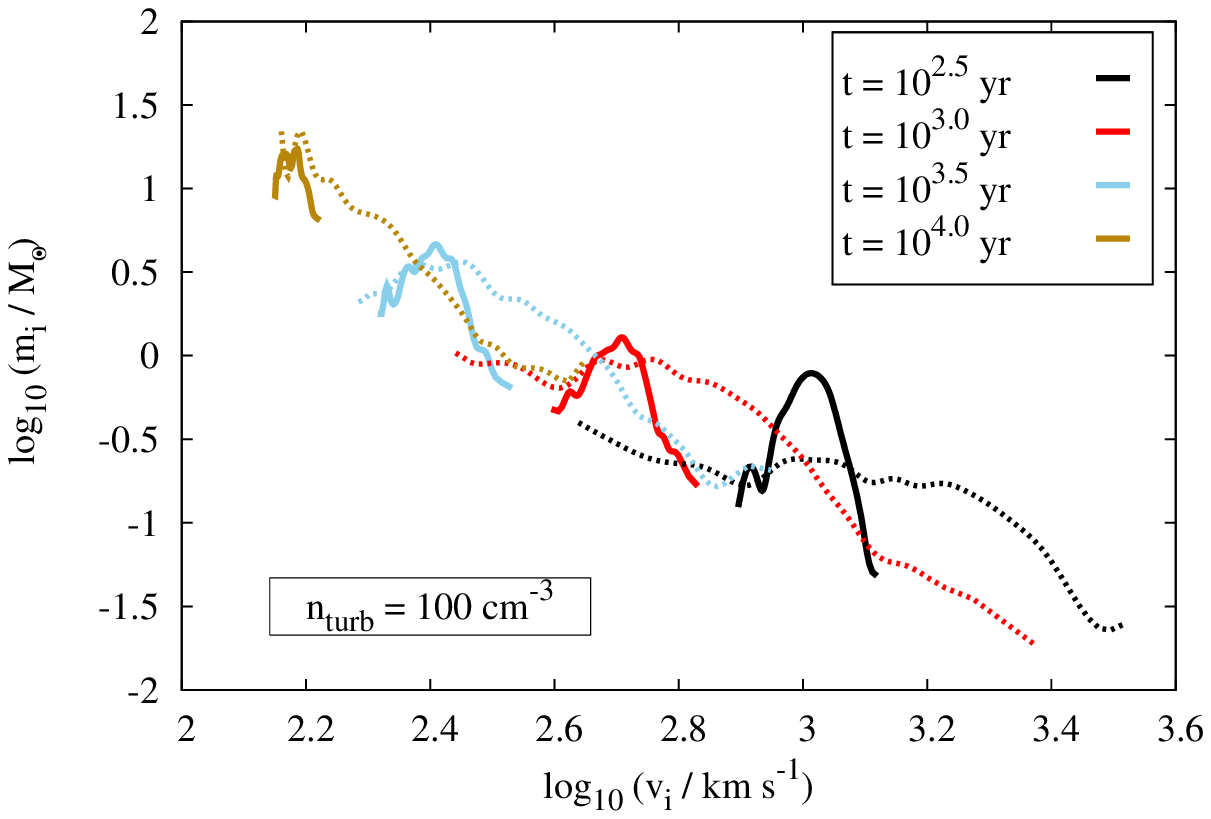}
\caption{Evolution of the mass-velocity distribution at times between $10^{2.5}$ yr and $10^{4.5}$ yr with different turbulent Mach numbers of  $\mathcal{M}$ = 1  (solid lines) and  $\mathcal{M}$ =100 (dashed lines). \textit{Top panel:} Low density environment with a mean ambient density $n_{_{0,\rm turb}}$ = 1 $\mathrm{cm^{-3}}$. \textit{Bottom panel:} Ambient medium with a density $n_{_{0,\rm turb}}$ = 100 $\mathrm{cm^{-3}}$.}
\label{fig:massvel}
\end{figure}

The SN model assumes that the swept-up ambient material is condensed into a small volume at the shock front \citep{klein94}. The density profile inside the SNR can be neglected as the mass is only a small fraction of the total mass. We show the distribution of the shock velocity and the swept-up mass to mean densities $n_{_{0,\rm turb}}$ of 1 $\mathrm{cm^{-3}}$ (Fig. \ref{fig:massvel}, top panel) and 100 $\mathrm{cm^{-3}}$ (Fig. \ref{fig:massvel}, bottom panel) with turbulent Mach numbers of 1 and 10 both with $N_{_{\mathrm{cones}}}$ = 384. The distributions are evaluated at fixed times between $t = 10^{2.5}$ yr and $t = 10^{4.5}$ yr. In dense environments ($n_{_{0,\rm turb}}$ = 100 $\mathrm{cm^{-3}}$) the simulations terminate earlier, explaining why in Fig. \ref{fig:massvel} (bottom panel) the distributions at $t = 10^{4.5}$ yr are missing.

As expected, the swept-up mass continuously increases during the decelerating expansion of the SNR. At $10^{2.5}$ yr the swept-up mass in a low density and low-$\mathcal{M}$-turbulence environment ($\mathcal{M} =1$, $n_{_{0,\rm turb}}$ = 1 $\mathrm{cm^{-3}}$) is 6.5 $\rm M_{\odot}$. For the case of $n_{_{0,\rm turb}}$ = 100 $\mathrm{cm^{-3}}$ the swept-up mass is 29.8 $\rm M_{\odot}$. In general higher-$\mathcal{M}-$turbulence results in lower swept-up masses, by 12 per cent in low- and 24 per cent in high-density environments. At $10^{4}$ yr the swept-up masses have increased to 280 $\rm M_{\odot}$ and 1279 $\rm M_{\odot}$ in the low- and high-density ambient medium. At this time the SNR evolution in the latter case has almost reached the end of the PDS, whereas in the first medium the PDS lasts longer, until $\sim 10^{5}$ yr.

The mean velocity at $t = 10^{2.5}$ yr is 2569 $\rm km\,s^{-1}$ in the low density environment. High-$\mathcal{M}$-turbulence increases the value to 3096 $\rm km\,s^{-1}$. The SNR slows down by $\sim$ 50 per cent in high density structures with $n_{_{0,\rm turb}}$ = 100 $\mathrm{cm^{-3}}$. Typically, at each plotted time the mean velocity decreases by $\sim$ 50 per cent compared with the previous time. At $t = 10^{4}$, the velocities have dropped to 323 $\rm km\,s^{-1}$ in low density structures with  $n_{_{0,\rm turb}}$ = 1 $\mathrm{cm^{-3}}$ and trans-sonic turbulence. In high density environment the mean velocity is 151 $\rm km\,s^{-1}$. 

At the end of the simulations, the distributions within an environment with trans-sonic turbulence cover a small velocity range. High-$\mathcal{M}$-turbulence broadens the mass-(shock-) velocity distribution and therefore, a small fraction of the swept-up mass remains at high velocities.

Similar behaviour is found in numerical simulations by \citet{walch14a}. At 0.2 Myr the velocity distribution in a dense ($n_{_{0,\rm turb}}$ = 100 $\mathrm{cm^{-3}}$) fractal environment shows that about 2 per cent of a cloud mass of $10^{5}$ $\rm M_{\odot}$ are accelerated to velocities larger than $\sim$ 20 $\rm km\,s^{-1}$.

 \section{Summary and discussion}
  \label{section7}
We present a fast model to follow the evolution of SN blast waves in their momentum generating phases (ST, TR and PDS phase). We test the model for homogeneous and power-law density distributions and extend it to the evolution of SNR in wind-blown bubbles and a turbulent ISM. Previous analytic work is combined in our SN model and extended by the inclusion of a cooling function, a detailed treatment of the thermal energy, and a transition phase between the adiabatic and radiative phase.

The main results are summarized in the following:
\begin{itemize}
\item We recover recent numerical results \citep[e.g.][]{kim14, martizzi14, li15}of a single SN in a homogeneous medium as well as the analytic Sedov-Taylor solution. The final momentum for a density range between 1$-$ 100 $\mathrm{cm^{-3}}$ is $\sim$ 13 $-$ 31 p$_{_{0}}$ (p$_{_{0}}$ = 14181 M$_{_{\odot}}$ km s$^{-1}$). We obtain reliable values for the radial momentum, the expansion radius and the thermal energy with small computational effort of a few seconds. The results depend solely on the ambient density.

\item In ambient media with a power-law density distribution and a surrounding density floor, the final momentum clearly exceeds the homogeneous results by at most a factor of 2. This is independent of the central density and is controlled by the density of the density floor. The inner power-law part has minor effect. 

\item The momentum input of SNR in wind-blown bubbles depend on the initial ambient medium. Low initial temperatures result in dense shells, where the incoming SN shock cools efficiently. The momentum input is only $\sim$ 3 p$_{_{0}}$. Higher temperatures of the initial ambient medium delay the radiative cooling in the wind-blown shell. The momentum input increases by a factor up to 10. A high density inside the bubble and a small difference between the swept-up mass and the mass of the wind-blown shell show a behaviour that is comparable to a uniform medium with that bubble density.

\item We use the SN model to approximate the lower limit of momentum input in turbulent ambient media. To do this we randomly sample densities from a log-normal density distribution with a given dispersion which is related to the Mach number in the turbulent gas. For low turbulent Mach numbers ($\mathcal{M}$ $\sim$ 1) the momentum input is very similar to homogeneous media ($\sim$ 13 $-$ 31 p$_{_{0}}$). We obtain the largest momentum input in turbulent media with $\mathcal{M}$ $\sim$ 100 by as much as a factor of 2 in a low density environment ($n_{_{0,\rm turb}}$ = 0.1 $\mathrm{cm^{-3}}$). We have parametrised the momentum input as a function of Mach number and average environmental density as follows:
\begin{multline}
p_{_{\rm turb}}/\mathrm{p_{_{0}}}\ =23.07\, (n_{_{0,\rm turb}}/ 1\,\mathrm{cm^{-3}})^{-0.12}\\ + 0.82 (\ln(1+b^{2}\mathcal{M}^{2}))^{1.49} (n_{_{0,\rm turb}}/ 1\ \mathrm{cm^{-3}})^{-0.17}.
\end{multline}
 Under the assumption of a neutral ambient medium we find values comparable to recent numerical simulations \citep[e.g.][]{kim14, martizzi14, walch14a}

\item The model is computational cheap and can be used for a variety of parameters. This model is an accurate alternative to recent SN sub-grid models.

\end{itemize}

\section*{Acknowledgments}
We thank J. P. Ostriker for the useful suggestions and discussion, which added significantly to the presented paper. SH, SW and DS acknowledge the support by the Bonn-Cologne Graduate School for physics and astronomy as well as the SFB 956 on the 'Conditions and impact of star formation'. JM acknowledges funding from a Royal Society - Science Foundation Ireland University Research Fellowship. We acknowledge the support by the DFG Priority Program 1573 'The physics of the interstellar medium'. We thank Anika Schmiedeke for her help of fitting with MAGIX and Thomas M\"oller for providing this tool. We thank the anonymous referees for constructive input.

\bibliographystyle{aa}
\bibliography{astro}

\appendix
\label{appendix}

\bsp

\label{lastpage}

\end{document}